\documentclass[11pt]{article}

\usepackage{mathrsfs,amsmath,amsbsy,amssymb,mcite,empheq,cite}

\def\be{\begin{equation}}
\def\ee{\end{equation}}
\def\bea{\begin{eqnarray}}
\def\eea{\end{eqnarray}}

\DeclareTextFontCommand{\textwasy}{\wasyfamily}
\def \wasyfamily{\fontencoding{U}\fontfamily{wasy}\selectfont}
\def \thorn{{\wasyfamily\char105}}
\DeclareTextCommand{\dh}{OT1}{{\wasyfamily\char107}}
\newcommand{\tho}{{\textrm\thorn}}
\renewcommand{\eth}{{\textrm{\dh}}}
\newcommand{\thop}{\tho'}

\newcommand{\Phia}{\Phi^\mathrm{A}} % misalignment of lower indices on \Phis and \Phia

\newcommand{\mbar}{{\bar{m}}}
\newcommand{\zbar}{{\bar{z}}}

\renewcommand{\d}{\mathrm{d}}

\title{Twisting algebraically special solutions in five dimensions} 

\author{Gabriel Bernardi de Freitas, Mahdi Godazgar and Harvey S. Reall\\[3mm] {\footnotesize Department of Applied Mathematics and Theoretical Physics, University of Cambridge}\\ {\footnotesize Centre for Mathematical Sciences, Wilberforce Road, Cambridge CB3 0WA, UK.}\\[1mm] {\footnotesize G.B.Freitas@damtp.cam.ac.uk, M.M.Godazgar@damtp.cam.ac.uk, H.S.Reall@damtp.cam.ac.uk}}

\date\today

\begin{document}

\maketitle

\begin{abstract}
 We determine the general form of the solutions of the five-dimensional vacuum Einstein equations with cosmological constant for which (i) the Weyl tensor is everywhere type II or more special in the null alignment classification of Coley \emph{et al.}, and (ii) the $3 \times 3$ matrix encoding the expansion, shear and twist of the aligned null direction has rank 2. The dependence of the solution on 2 coordinates is determined explicitly, so the Einstein equation reduces to PDEs in the 3 remaining coordinates, just as for four-dimensional algebraically special solutions. The solutions fall into several families. One of these consists of warped products of four-dimensional algebraically special solutions. The others are new.
\end{abstract}

\section{Introduction and Summary}

\subsection{Introduction}

There is a long tradition of classifying spacetimes according to the algebraic properties of their Weyl tensor. In four dimensions, this is the Petrov classification. A spacetime is said to be algebraically special if it is everywhere of Petrov type II or more special. This is a useful assumption to make when searching for solutions of the vacuum Einstein equation
\be
 R_{ab} = \Lambda g_{ab},
\label{Einstein:eqn}
\ee
which will be the subject of this paper. It leads to a considerable simplification that allows one to integrate the Einstein equations completely in one of the coordinates and reduces the problem to a small set of PDEs in the other three \cite{Kerr:1963ud, Debney:1969zz, RRZ, Trim:1974, exactsolutions, Timofeev1996, KaigorodovTimofeev1996}. The general solution of these PDEs is not known, but some important cases are the Kerr solution and the $C$-metric.

In practice, the usefulness of this approach is closely connected with the \emph{Goldberg-Sachs theorem}, which states that an \emph{Einstein spacetime}, i.e.\ a solution of \eqref{Einstein:eqn}, is algebraically special if, and only if, it admits a geodesic, shear-free null congruence. The null vector field $\ell$ tangent to this congruence is a \emph{repeated principal null direction} (PND) of the Weyl tensor, satisfying
\be
 \ell^b  \ell_{[e} C_{a]b[cd} \ell_{f]} = 0.
\label{rep:pnd}
\ee
There is then a natural coordinate on such spacetimes, an affine parameter $r$ along each geodesic tangent to $\ell$. The restrictions imposed by the algebraically special condition, including the Goldberg-Sachs theorem, then allow one to integrate the Einstein equations \eqref{Einstein:eqn} in this coordinate and determine the complete $r$-dependence of the metric \cite{exactsolutions, KaigorodovTimofeev1996}.

In higher dimensions, an algebraic classification of the Weyl tensor was proposed by Coley, Milson, Pravda and Pravdova \cite{cmpp}. This classification is based on null alignment and generalises the concept of a repeated PND to higher dimensions. A higher-dimensional spacetime is said to be algebraically special if it admits a \emph{multiple Weyl aligned null direction} (WAND), which is a null vector field $\ell$ satisfying the same eqn.~\eqref{rep:pnd} as in the four-dimensional case \cite{ortaggio}.

To exploit the algebraically special condition to solve \eqref{Einstein:eqn} in higher dimensions would require a higher-dimensional generalisation of the four-dimensional Goldberg-Sachs theorem. The ``geodesic part" of this theorem was addressed in Ref.\ \cite{geodesic}. There, it was shown that, although multiple WANDs need not be geodesic, a spacetime admits a multiple WAND if, and only if, it also admits a \emph{geodesic} multiple WAND. Therefore, there is no loss in generality by taking $\ell$ satisfying \eqref{rep:pnd} to be tangent to null geodesics.

The generalisation of the shear-free condition is less straightforward. For a null vector field $\ell$ in an $n$-dimensional spacetime, let $m_i$ be a set of $n-2$ orthonormal, spacelike vectors transverse to $\ell$, i.e.\ $\ell \cdot m_i = 0, m_i \cdot m_j = \delta_{ij}$. The \emph{optical matrix} of $\ell$ is defined by the $(n-2) \times (n-2)$ matrix
\be
 \rho_{ij} \equiv m_i^a m_j^b \nabla_b \ell_a.
\ee
The expansion, shear and twist of $\ell$ then correspond respectively to the trace, symmetric traceless part, and antisymmetric part of $\rho_{ij}$. Unlike the four-dimensional Kerr solution, it is known that rotating black holes in higher dimensions admit multiple WANDs that are shearing \cite{myersperry, frolov,typed}. Therefore the shear-free condition does not extend in a straightforward way to higher dimensions. It turns out that the correct way of generalizing this condition takes the form of algebraic restrictions on the form of $\rho_{ij}$ when $\ell$ is a multiple WAND. General results restricting the optical matrix of a multiple WAND in arbitrary dimensions are not known except in some special cases, e.g.\ type III and N spacetimes \cite{bianchi} and non-twisting multiple WANDs \cite{nontwisting}. However, in five dimensions, which will be the subject of this paper, Ref.\ \cite{gspaper} has determined the allowed forms of $\rho_{ij}$.

In five dimensions, Ref.\ \cite{gspaper} showed that the spatial vectors $m_i$ ($i=2,3,4$) can be chosen so that the optical matrix $\rho_{ij}$ of a multiple WAND $\ell$ takes one of the following forms:
\be \label{rho}
 b\left( \begin{array}{ccc} 1 & a & 0 \\ -a & 1 & 0 \\ 0 & 0 & 1+a^2 \end{array} \right), \qquad b\left( \begin{array}{ccc} 1 & a & 0 \\ -a & 1 & 0 \\ 0 & 0 & 0 \end{array} \right), \qquad b\left( \begin{array}{ccc} 1 & a & 0 \\ -a & -a^2 & 0 \\ 0 & 0 & 0 \end{array} \right)
\ee
for spacetime functions $a,b$. For $b \ne 0$, these matrices have ranks 3, 2 and 1, respectively. The case $b = 0$ corresponds to the Kundt family \cite{kundt1, kundt2, kundt3}. In our previous work \cite{5drank3}, we have considered the rank three case and shown that the only such solution is the five-dimensional Kerr-de Sitter spacetime or some analytic continuation thereof. All such spacetimes are type D. The case of a rank 1 optical matrix has been investigated by Wylleman. He has proved \cite{wylleman1} that the class of such solutions coincides with the class of solutions that admit a non-geodesic multiple WAND (see also \cite{gspaper}), which were completely classified in Ref.\ \cite{geodesic}. All such solutions are type D. 

In this paper, we complete the classification of algebraically special Einstein spacetimes in five dimensions by considering the rank 2 case,
\be
 \rho_{ij} = b \left(
\begin{array}{ccc}
1 & a & 0 \\
-a & 1 & 0 \\
0 & 0 & 0
\end{array}
\right).
\label{rho1}
\ee
When $a = 0$, the null vector field $\ell$ is non-twisting (and geodesic) and therefore \emph{hypersurface-orthogonal}. All solutions in this case were determined in Ref.\ \cite{hyporthog}.

The main aim of this paper is to generalise the results of Ref.\ \cite{hyporthog} to the twisting case $a \ne 0$~\footnote{Our treatment of the general twisting case will be such that our results also include the non-twisting $a=0$ case.  In particular, we rederive here the solutions presented in Ref.\ \cite{hyporthog}.}. 
We follow the same method as in our previous work \cite{5drank3}. We work in a null frame $\{\ell,n,m_i\}$ where $\ell$ is the multiple WAND, the $m_i$ are chosen so that $\rho_{ij}$ takes the form \eqref{rho1} and $n$ is another null vector field orthogonal to $m_i$. In this basis the curvature components can be determined by the ``Newman-Penrose" equations of Refs.\ \cite{ricci, ghp}, constrained by the Bianchi identity. 

Since $\ell$ can be taken to be geodesic, we choose an affine parameter $r$ along these geodesics as one of the coordinates. Then $\ell$ acts as a partial derivative with respect to $r$. In section \ref{sec:rbasis}, we integrate some of the Newman-Penrose equations, some components of the Bianchi identity, and commutators involving the basis vectors, to determine the complete $r$-dependence of the basis vectors and hence also the metric. 

In section \ref{sec:constraint}, we examine the remaining equations constraining the curvature and basis vectors. We then show in section \ref{sec:coords} how to introduce local coordinates adapted to the properties of these solutions. With this method, we are able to determine fully the dependence of the metric on another coordinate $x$. The Einstein equation then reduces to PDEs in 3 coordinates, just as in four dimensions. 

Before we summarise our results, it is helpful to introduce notation by reviewing the general form of algebraically special solutions in four dimensions.  

\subsection{Algebraically special solutions in four dimensions}
\label{sec:4dalgs}

We will consider four-dimensional solutions of \eqref{Einstein:eqn}. For future convenience we will denote the 4d cosmological constant by $\lambda$ instead of $\Lambda$. In four dimensions, the optical matrix of a repeated PND can have rank 0 or 2. Rank 0 defines the Kundt family of solutions. We will consider the rank 2 case. Coordinates $(u,R,z,\bar{z})$ can be introduced so that the solution is written in terms of a complex function $L(u,z,\zbar)$ and real functions $P(u,z,\zbar)$, $m(u,z,\zbar)$ and $M(u,z,\zbar)$. The metric takes the form \cite{exactsolutions, KaigorodovTimofeev1996}:
\be
\label{4dalgspec}
 g_{(4)} = 2 \ell \left( \d R + U \d z + \bar{U} \d \zbar - \mathcal{H} \ell \right) + \frac{2 (R^2 + \Sigma^2)}{P^2} \d z \d \zbar,
\ee
where 
\begin{equation} \label{def:ell}
 \ell = -(\d u +L \d z + \bar{L} \d \bar{z}),
\end{equation}
\be
 2 i\, \Sigma = P^2 (\bar{\partial} L - \partial \bar{L}), 
 \label{def:Sigma}
\ee
\be
\label{def:U}
 U = i \partial \Sigma - \left( R + i \Sigma \right) \partial_u L,
\ee
\be
 \mathcal{H} = \frac{P^2}{2} (\partial \bar{S} + \bar{\partial} S) - \frac{\lambda}{6} R^2 - R \partial_u \ln P - \frac{m R + M \Sigma}{R^2 + \Sigma^2}
\ee
with
\begin{align} \label{def:S}
 S = \partial \ln P - \partial_u L.
\end{align}
In the above expressions we have used derivative operators $\partial,\bar{\partial}$ defined by
\begin{equation}
\label{ddef}
 \partial \equiv \partial_z - L \partial_u, \qquad \bar{\partial} \equiv \partial_{\bar{z}} - \bar{L} \partial_u.
\end{equation}
These satisfy the following commutator relations:
\begin{equation} \label{comm}
[\partial, \bar{\partial}] = (\bar{\partial} L - \partial \bar{L})\ \partial_u, \qquad [\partial, \partial_u] = \partial_u L\ \partial_u.
\end{equation}
The repeated principal null direction is given by raising an index on the 1-form $\ell$:
\be
 \ell = \frac{\partial}{\partial R}.
\ee
As a vector field, $\ell$ is geodesic and shear-free, in agreement with the Goldberg-Sachs theorem. The coordinate
$R$ is an affine parameter along the geodesics with tangent $\ell$. Notice that the $R$-dependence of the metric is fully determined. 

If we define a complex function $k$ by $m+iM = -k$ then the Einstein equation reduces to the following set of PDEs
\begin{gather}
 \mathrm{Im} (k) = - P^2 \mathrm{Re} (\partial \bar{\partial} \Sigma - 2 \partial_u \bar{L} \partial \Sigma - \Sigma \partial_u \partial \bar{L}) - 2 \Sigma P^2 \mathrm{Re} (\partial \bar{S}) + \frac{4 \lambda \Sigma^3}{3}, \notag \\[2mm]
 (\partial-3 \partial_u L) k =0, \notag \\[2mm]
 - \partial_u \left( \frac{k}{P^3} \right) = P \left( \partial + 2S \right) \partial (\bar{S}^2 + \bar{\partial} \bar{S}) - \lambda P \left( \partial + 2S \right) \left[ \frac{1}{P^2} (\bar{\partial} \Sigma^2 - 2 \Sigma^2 \partial_u \bar{L}) \right].
\end{gather}
Note that the first equation above fixes the function $M$. 

\subsection{Summary of Results}

\label{sec:introsummary}

Any solution of the 5d vacuum Einstein eqn.\ \eqref{Einstein:eqn} which admits a multiple WAND with an optical matrix of rank 2 must belong to one of the classes described below. Note that the derivation of these results presented in the main part of this paper does not assume that the multiple WAND is twisting, except when the distinction is important in certain cases. Therefore, the results written below include all the non-twisting solutions of Ref.\ \cite{hyporthog}. All solutions are written in terms of certain functions $L,Y,\Sigma,j,k,P,F$.  In all cases, the multiple WAND is non-twisting (i.e.\ hypersurface-orthogonal) if, and only if, $\Sigma=0$. When this condition is satisfied one can use a residual coordinate freedom to set $L=0$.

\paragraph{Class 1.}
Warped product solutions the form
\be
\label{warped}
ds^2 = \d x^2 + F(x)^2 g_{(4)},
\ee
where $g_{(4)}$ is a four-dimensional (non-Kundt) algebraically special solution with cosmological constant $\lambda$, i.e.\ a metric of the form \eqref{4dalgspec}. The warp factor $F(x)$ satisfies
\begin{equation}
 F'' + \frac{\Lambda}{4} F = 0,
 \label{Feq1}
\end{equation}
with first integral
\begin{equation}
 F'^2+ \frac{\Lambda}{4}F^2 = \frac{\lambda}{3}.
\label{eqn:F}
\end{equation}
Of course these equations are easily solved to determine $F(x)$ explicitly for any given $\Lambda$ and $\lambda$. The multiple WAND is $\partial/\partial R$. 

These solutions are generically of type II. The condition for the solution to be type III (or more special) is $k = 0$ (with $k$ defined above in section \ref{sec:4dalgs}), which is the same as the condition for $g_{(4)}$ to be type III or more special \cite{exactsolutions, KaigorodovTimofeev1996}. 

If $g_{(4)}$ is of type D then so is the 5d metric. All type D vacuum solutions in 5d have been classified in Ref.~\cite{wylleman}. The classification reveals that any type D solution for which one (or both) of the two multiple WANDs has a rank 2 optical matrix can be written as in \eqref{warped} where $g_{(4)}$ is type D \cite{wylleman}.

\paragraph{Class 2.} For these solutions, either $\Lambda=0$ or $\Lambda<0$ and we treat these two possibilities separately.
\subparagraph{2.1} $\Lambda = 0.$  
We can introduce coordinates $(u,r,x,z,\bar{z})$ so that the multiple WAND is given by
\begin{equation}
 \ell = \frac{\partial}{\partial r}.
\end{equation}
The solution is given in terms of complex functions $L(u,z,\zbar)$, $Y(u,z,\zbar)$ and real functions $J(u,z,\zbar)$, $P(u,z,\zbar)$, $m(u,z,\zbar)$ and $M(u,z,\zbar)$. The metric takes the form
\begin{equation}
ds^2 = \left( \d x + Y \d z + \bar{Y} \d \bar{z} - J \ell \right)^2 + 2 \ell \left( \d r + U \d z + \bar{U} \d \zbar - \mathcal{H} \ell \right) + \frac{2 (r^2 + \Sigma^2)}{P^2} \d z \d \zbar,
\end{equation}
where $\ell,$ $\Sigma$ and $U$ are defined as above in eqns.~\eqref{def:ell}, \eqref{def:Sigma}, \eqref{def:U} with $R=r$ and real $J$ satisfies 
\be
   \Sigma J = \frac{P^2}{2} i (\bar{\partial} Y - \partial \bar{Y}).
  \label{def:J}
\ee
Note that when $\Sigma \ne 0$, this determines $J$ in terms of the other functions.  The function $\mathcal{H}$ is defined by
\be 
  \mathcal{H} = \frac{P^2}{2} (\partial \bar{S} + \bar{\partial} S) - r \partial_u \ln P - \frac{(m-x j) r + M \Sigma}{r^2 + \Sigma^2},
\ee
where $S$ is defined in eqn.\ \eqref{def:S},  $\partial$ defined by \eqref{ddef}, and the  real function $j$ is defined by\footnote{Although it is not obvious as defined that Im$(j)$=0, this can be verified by a direct calculation. For $\Sigma=0$, this follows from the fact that a gauge can be chosen such that $L=0$ and from eqn.\ \eqref{def:J}, we have that $\partial_z \bar{Y} = \partial_{\bar{z}}Y$.  For $\Sigma\neq 0$, we show this by using the commutator relations \eqref{comm}, the definition of $\Sigma$ given in \eqref{def:Sigma} and the definition of $J$, eqn.~\eqref{def:J}.}
\begin{equation} \label{def:j}
 j = P^2 (\partial - \partial_u L) (\bar{\partial} J - J \partial_u \bar{L} - \partial_u \bar{Y}).
\end{equation}
This quantity must satisfy the following equations
\begin{equation}
 \partial_u \left(\frac{j}{P^3} \right) = 0, \qquad \left( \partial - 3 \partial_u L \right) j = 0.
\end{equation}
Note that the only $x$-dependence in the metric is via a term in $\mathcal{H}$ that is linear in $x$.

The quantity $k\equiv -(m+ iM)$ must satisfy the following constraints
\begin{gather}
 \mathrm{Im} (k) = - P^2 \mathrm{Re} (\partial \bar{\partial} \Sigma - 2 \partial_u \bar{L} \partial \Sigma - \Sigma \partial_u \partial \bar{L}) - 2 \Sigma P^2 \mathrm{Re} (\partial \bar{S}), \notag \\[2mm]
 \partial_u \left( \frac{k}{P^3} \right) =  J \frac{j}{P^3} + \frac{1}{2P} |\partial J - J \partial_u L - \partial_u Y|^2 - P \left( \partial + 2S \right) \partial (\bar{S}^2 + \bar{\partial} \bar{S}), \notag \\[2mm]
 \left( \partial - 3 \partial_u L \right) k = Y j.
\end{gather}
From the above equations one can derive the following
\begin{gather}
j \partial_u \left[ P (\partial \bar{L} - \bar{\partial}L) \right] = 0,  \qquad j \partial_u S = 0.
\end{gather}
Therefore this subclass naturally divides further into two subsubclasses. 

\subparagraph{2.1.1} $j \ne 0$. In this case we have the additional equations
\begin{equation} \label{2.1.1con}
\partial_u \left[ P (\partial \bar{L} - \bar{\partial}L) \right] = \partial_u S = 0.
\end{equation}
Such solutions are strictly type II.

\subparagraph{2.1.2} $j = 0.$ In this case, the metric is $x$-independent and can be written in the Kaluza-Klein form
\be
\label{KK}
 ds^2 = (\d x + \mathcal{A})^2 + g_{(4)}
\ee
with Kaluza-Klein gauge field
\be
 \mathcal{A} = Y \d z + \bar{Y} \d \zbar - J \ell.
\ee
Note that the dilaton (Kaluza-Klein scalar field) is constant. 

The pair $(g_{(4)},\mathcal{A})$ is a solution of the equations for four-dimensional (non-Kundt) algebraically special solutions of Einstein-Maxwell theory with vanishing cosmological constant and an aligned null Maxwell field \cite{exactsolutions}. Conversely, any such 4d solution can be oxidised to give a 5d algebraically special solution of the form (\ref{KK}).

If a solution in this class is type III (or more special) then $d{\cal A}=0$, so $\mathcal{A}$ can be eliminated by a transformation  $x \rightarrow x + f(u,z,\zbar)$. Therefore, any such solution is simply a product of a flat direction with a 4d Ricci flat solution of type III (or more special) and hence also belongs to Class 1.

\subparagraph{2.2} $\Lambda < 0.$ Define
\begin{equation}
 l = \sqrt{- 4/\Lambda}
\end{equation}
We can introduce coordinates $(u,R,x,z,\bar{z})$ so that the multiple WAND is $\partial/\partial R$. The solution is given in terms of complex functions $L(u,z,\zbar)$, $Y(u,z,\zbar)$ and real functions $J(u,z,\zbar)$, $P(u,z,\zbar)$, $m(u,z,\zbar)$ and $M(u,z,\zbar)$. The metric is
\begin{equation} \label{met:2.2}
ds^2 = \left( \d x + Y \d z + \bar{Y} \d \bar{z} - J \ell \right)^2 +e^{2x/l} \left[ 2 \ell \left( \d R + U \d z + \bar{U} \d \zbar - \mathcal{H} \ell \right) + \frac{2 (R^2 + \Sigma^2)}{P^2} \d z \d \zbar \right],
\end{equation}
where the 1-form $\ell$ is defined by \eqref{def:ell}, $\Sigma$ is defined in \eqref{def:Sigma} and
\begin{gather}
 U = i \partial \Sigma - (R + i \Sigma) \left( \partial_u L + \frac{2 Y}{l} \right), \notag \\[2mm]
 \mathcal{H} = \frac{P^2}{2} \left[ \partial \left( \bar{S} - \frac{\bar{Y}}{l} \right) + \bar{\partial} \left( S - \frac{Y}{l} \right) \right] - R \left( \partial_u \ln P + \frac{J}{l} \right) - \frac{\left[ m + \frac{l j}{2} e^{-2x/l} \right] R + M \Sigma}{R^2 + \Sigma^2},
 \label{2.2:UH}
\end{gather}
where $S$, $\partial$ are defined as in eqns.~\eqref{def:S}, \eqref{ddef} and (real) $j$ is given by \eqref{def:j}. Note that the only $x$-dependence in the metric arises from the $e^{2x/l}$ in (\ref{met:2.2}) and the $e^{-2x/l}$ in $\mathcal{H}$. 

The solution must satisfy eqn.~\eqref{def:J}. This equation determines $J$ if $\Sigma \ne 0$. The function $j$ must satisfy 
\begin{equation}
 \frac{l}{2} \partial_u \left( \frac{j}{P^3} \right) + \frac{J j}{2 P^3} + \frac{1}{2P} |\partial J - J \partial_u L - \partial_u Y|^2 = 0, \qquad \partial j = 3 j \partial_u L + \frac{2 Y j}{l},
\end{equation}
while $k=-(m+iM)$ satisfies
\begin{gather}
 \mathrm{Im} (k) = - P^2 \mathrm{Re} \left[ \partial \bar{\partial} \Sigma - 2 \partial_u \bar{L} \partial \Sigma - \Sigma \partial_u \partial \bar{L} + 2 \Sigma \partial \bar{S} - \frac{4}{l} \partial (\bar{Y} \Sigma) + \frac{4}{l} \bar{Y} \Sigma \partial_u L + \frac{4 |Y|^2 \Sigma}{l^2} \right], \notag \\[2mm]
 \partial_u \left( \frac{k}{P^3} \right) - \frac{J k}{l P^3} + i \frac{\Sigma j}{l P^3} + \frac{2}{l P} i \left[ \partial \Sigma - \Sigma \left( \partial_u L + \frac{2 Y}{l} \right) \right] \left( \bar{\partial} J - J \partial_u \bar{L} - \partial_u \bar{Y} \right) \hspace{20mm} \nonumber \\
   \hspace{50mm} + P \left[ \partial + 2 \left( S - \frac{Y}{l} \right) \right]  \partial \left[ \left( \bar{S} - \frac{\bar{Y}}{l} \right)^2 + \bar{\partial} \left( \bar{S} - \frac{\bar{Y}}{l} \right) \right] = 0, \notag \\[2mm]
   \partial k = 3 k \partial_u L + \frac{4 Y k}{l}.
\end{gather}
These solutions are generically of type II. If a solution in this class is type III (or more special) then a coordinate transformation can be used to bring it to the warped product form of Class 1.

\paragraph{Class 3}

We can introduce coordinates $(u,R,x,z,\bar{z})$ so that the multiple WAND is $\partial/\partial R$. 
The solution is given in terms of complex functions $L(u,z,\zbar)$, $Y(u,z,\zbar)$ and real functions $J(u,z,\zbar)$, $P(u,z,\zbar)$. The metric is
\be
 ds^2 = (\d x + Y \d z + \bar{Y} \d \zbar - J \ell)^2 + F^2 \left[ 2 \ell \left( \d R + U \d z + \bar{U} \d \zbar - \mathcal{H} \ell \right) + \frac{2 (R^2 + \Sigma^2)}{P^2} \d z \d \zbar \right],
\ee
where (real) $F(x)$ is determined explicitly by solving \eqref{Feq1} and \eqref{eqn:F} with $\lambda \ne 0$ ($\lambda=0$ gives Class 2). The 1-form $\ell$ is defined by \eqref{def:ell} and $\Sigma$ is defined by \eqref{def:Sigma}. The functions $U$ and $\mathcal{H}$ are given by
\begin{gather}
 U = i \partial \Sigma - (R + i \Sigma) \left( \partial_u L + \frac{2 Y F'}{F} \right), \notag \\[2mm]
 \mathcal{H} = \frac{P^2}{2} (\partial \bar{S} + \bar{\partial} S) - \frac{P^2 F'}{2 F^2} (\partial \bar{Y} + \bar{\partial} Y) - \frac{\lambda P^2 |Y|^2}{3 F^2} - \frac{\lambda R^2}{6} - R \left( \partial_u \ln P + \frac{J F'}{F} \right),
\end{gather}
where $S$ is defined in \eqref{def:S} and $\partial$ in \eqref{ddef}. Note that the $x$-dependence is fully determined in terms of $F(x)$. 

The various functions in the metric must satisfy \eqref{def:J}. This equation determines $J$ if $\Sigma \ne 0$. The other equations are 
\begin{gather}
 (\partial - \partial_u L) \left( \partial J - J \partial_u L - \partial_u Y + \frac{4 i \lambda Y \Sigma}{3} \right) = 0, \qquad \mathrm{Re} \left[ \partial(\bar{Y}\Sigma) - \bar{Y} \Sigma \partial_u L \right] = 0 , \notag \\[2mm]
 P^2 \mathrm{Re} \left[ \partial \bar{\partial} \Sigma - 2 \partial_u \bar{L} \partial \Sigma - \Sigma \partial_u \partial \bar{L} + 2 \Sigma \partial \bar{S} \right] = \Lambda P^2 |Y|^2 \Sigma + \frac{4 \lambda \Sigma^3}{3}, \notag \\[2mm]
 \left| \partial J - J \partial_u L - \partial_u Y + \frac{4 i \lambda Y \Sigma}{3} \right|^2 
 + \frac{2 \lambda P^2}{3}|\partial Y + 2 Y S|^2 = 0, \notag \\[2mm]
  (\bar{\partial} + 2 \bar{S})\, {\Xi}_1 - \frac{\Lambda}{2}\, \bar{Y}\, {\Xi}_2 = 0, \qquad
 (\bar{\partial} + 2 \bar{S})\, {\Xi}_2 + 2\, \bar{Y}\, {\Xi}_1 = 0,
 \label{3:cons}
\end{gather}
where
\begin{gather}
 \Xi_1 = \bar{\partial} \left( S^2 + \partial S - \frac{\Lambda Y^2}{4} \right) - \frac{\lambda}{P^2} \left( \partial \Sigma^2 - 2 \Sigma^2 \partial_u L \right), \\[2mm]
 \Xi_2 =  \bar{\partial} (\partial Y + 2 Y S) + \frac{2i\, \Sigma}{P^2} \left( \partial J - J \partial_u L - \partial_u Y + \frac{4 i \lambda Y \Sigma}{3} \right).
\end{gather}

Type III (or more special) solutions are characterised by
\be
\label{case3type3cond}
 \partial J - J \partial_u L - \partial_u Y + \frac{4 i \lambda Y \Sigma}{3} = 0.
\ee
In fact it can be shown that if this condition is satisfied and $Y \ne 0$ then the solution is even more special, i.e., type N or conformally flat. Note that the equation on the third line of \eqref{3:cons} implies that \eqref{case3type3cond} is always satisfied when $\lambda > 0$. Hence, all solutions in this class with $\lambda>0$ are type III (or more special) and solutions with $\lambda>0$, $Y \ne 0$ are type N or conformally flat.

\section{Determining the $r$-dependence} \label{sec:rbasis}

\subsection{Notation}

We perform most calculations in a null basis. In Refs.\ \cite{bianchi,ricci} a higher-dimensional analogue of the Newman-Penrose formalism has been developed for calculations in such a basis. This is repackaged into a  higher-dimensional analogue of the Geroch-Held-Penrose (GHP) formalism in Ref.\ \cite{ghp}. We will follow the notation of Ref.\ \cite{ghp} for the connection components and Weyl tensor components. In particular, we refer the reader to eqns.~NP1--NP4, B1--B8 and C1--C3 of Ref.\ \cite{ghp}, which list all the Newman-Penrose and Bianchi equations satisfied by the connection and curvature components, as well as equations for the commutators of derivatives.

\subsection{Choice of basis}

Consider an Einstein spacetime, i.e.\ one satisfying \eqref{Einstein:eqn}, admitting a multiple WAND $\ell$ with a rank 2 optical matrix. Introduce a basis $\{\ell, n, m_i\}$, where $n$ (like $\ell$) is null and the $m_i$ ($i=2,3,4$) are spacelike and such that
\be
 \ell \cdot n = 1, \qquad m_i \cdot m_j = \delta_{ij},
\ee
with all other inner products zero. The multiple WAND condition for $\ell$ is equivalent to the vanishing of the positive boost weight components of the Weyl tensor, namely
\be
 \Omega_{ij} = 0, \qquad \Psi_{ijk} = 0.
\ee
As mentioned before, Ref.\ \cite{geodesic} showed that, without loss of generality, $\ell$ can be taken to be geodesic, hence we choose
\be
 \kappa_i = 0.
\ee
From \cite{gspaper} we know that the $m_i$ can be chosen so that the optical matrix takes the form \eqref{rho1}. 
As in our previous work \cite{5drank3}, it is convenient to combine the spatial basis vectors $m_2, m_3$ into the complex null vectors\footnote{A similar $2+2+1$ complex frame is independently employed in Ref.\ \cite{wylleman, wylleman1}.}
\be
 m_5 = \frac{m_2 + i m_3}{\sqrt{2}}, \qquad \mbar_5 = \frac{m_2 - i m_3}{\sqrt{2}}.
\ee
The optical matrix of $\ell$ is then
\be
  \rho_{ij} = b \left(
\begin{array}{ccc}
0 & 0 & 0 \\
0 & 0 & 1 - i a \\
0 & 1 + i a & 0
\end{array}
\right).
\label{rho2}
\ee
The structure of $\rho_{ij}$ given in \eqref{rho1} or \eqref{rho2} is preserved under null rotations about $\ell$ and spins in the 2--3 directions. We can then exploit these transformations and choose our basis so that some GHP scalars (and other objects) vanish. First note that we can rescale $\ell$ to be tangent to \emph{affinely parameterised} geodesics, hence $L_{10} = 0$. Next, consider eqn.\ NP1 of Ref.\ \cite{ghp}:
\be
 D \rho_{ij} + \stackrel{k}{M}_{i0} \rho_{kj} + \stackrel{k}{M}_{j0} \rho_{ik} = - \rho_{ik} \rho_{kj}.
\ee
Setting $ij = 45$ gives $\stackrel{4}{M}_{50} = 0$. Now, consider a spin in the 2--3 directions, which acts on $m_5, \mbar_5$ as follows:
\be
 m_5 \rightarrow e^{i \lambda} m_5, \qquad \mbar_5 \rightarrow e^{- i \lambda} \mbar_5
\ee
for some real function $\lambda$. The effect of such a spin on $\stackrel{i}{M}_{j0}$ is
\be
 \stackrel{4}{M}_{50} \rightarrow e^{i \lambda} \stackrel{4}{M}_{50}, \qquad \stackrel{5}{M}_{\bar{5}0} \rightarrow \stackrel{5}{M}_{\bar{5}0} + i D \lambda.
\ee
Hence, $\stackrel{4}{M}_{50}$ remains zero under this transformation.  An appropriate choice of $\lambda$ can then be used to set
\begin{equation}
 \stackrel{5}{M}_{\bar{5}0} = 0.
\end{equation}
Thus, we can find a basis such that
\begin{equation}
 \stackrel{i}{M}_{j0} = 0.
\end{equation}
Now consider a null rotation about $\ell$ with parameters $z_i$ \cite{ghp},
\be
\ell \longrightarrow \ell, \qquad n \longrightarrow n + z_i m_i - \frac{1}{2} z^2 \ell, \qquad m_i \longrightarrow m_i - z_i \ell,
\label{nullrot}
\ee
where $z^2 = z_i z_i$. Because $\kappa_i = 0$, both $\rho_{ij}$ and $\stackrel{i}{M}_{j0}$ remain unchanged under this transformation \cite{ghp}. However, $\tau_i$ transforms non-trivially \cite{ghp}
\be
 \tau_i \longrightarrow \tau_i + \rho_{ij} z_j.
\ee
We see that $\tau_4$ remains unchanged, but $\tau_5$ changes according to
\be
 \tau_5 \longrightarrow \tau_5 + b (1 - i a) z_5.
\ee
Because $b \neq 0$ ($b = 0$ is the Kundt case) and $a$ is real, we can choose $z_5$ to set $\tau_5 = 0$.  If we now look at eqn.\ NP2 of Ref.\ \cite{ghp},
\be
 D \tau_i = \rho_{ij} ( \tau'_j -\tau_j ),
\ee
$i = 5$ implies $\tau'_5 = 0$. Finally, consider a second null rotation \eqref{nullrot} about $\ell$ with parameters $z_4 \neq 0$, $z_5 = 0$. For general $z_i$, $\tau'_i$ changes as follows:
\be
 \tau'_i \longrightarrow \tau'_i + D z_i.
\ee
Therefore, with only $z_4$ non-vanishing, $\tau_5$ remains zero, while
\be
 \tau'_4 \longrightarrow \tau'_4 + D z_4.
\ee
We can then choose $z_4$ appropriately in order to set $\tau'_4 = 0$.

In summary, then, a basis can be chosen such that the following conditions hold:
\be
 L_{10} = 0, \qquad \stackrel{i}{M}_{j0} = 0, \qquad \kappa_i = 0, \qquad \tau_5 = 0, \qquad \tau'_i = 0.
\ee
This basis is \emph{parallelly transported} along geodesics with tangent $\ell$.

\subsection{Solving NP1 and NP2}

For the purpose of the following calculations, we introduce local coordinates as follows. Let $\Sigma$ be a hypersurface transverse to $\ell$ and introduce coordinates $x^\mu$ on $\Sigma$. Now assign coordinates $(r,x^\mu)$ to the point parameter distance $r$ along the integral curve of $\ell$ through the point on $\Sigma$ with coordinates $x^\mu$. We then have
\be
 \ell = \frac{\partial}{\partial r}.
\label{r-def}
\ee
Eqn.\ NP1 of Ref.\ \cite{ghp} in our basis is now
\be
 D \rho_{ij} = - \rho_{ik} \rho_{kj}.
\label{11g}
\ee
The only non-trivial component of this equation is $ij = 5\bar{5}$, which reads
\be
 D \rho_{5\bar{5}} = - (\rho_{5 \bar{5}})^2,
\ee
to which the solution is
\be
 \rho_{5\bar{5}} = \frac{1}{r - i \chi},
\ee
for some complex function $\chi$ which does not depend on $r$. The construction leading to eqn.\ \eqref{r-def} does not define $r$ uniquely; there is a freedom to shift it by a function of the $x^{\mu}$. This is equivalent to picking a different transverse hypersurface $\Sigma$. This freedom can then be used to make $\chi$ \emph{real}. In  summary, eqn.\ \eqref{11g} gives\footnote{We note that our result for $\rho_{ij}$ exhibits the same $r$-dependence as in Ref.\ \cite{KS-paper}.}
\be
 \rho_{ij} = \left(
\begin{array}{ccc}
0 & 0 & 0 \\
0 & 0 & \frac{1}{r - i \chi} \\
0 & \frac{1}{r + i \chi} & 0
\end{array}
\right),
\label{rho:r-dep}
\ee
where $\chi$ is a \emph{real} function of the coordinates $x^{\mu}$ only.

Similarly, eqn.\ NP2 of Ref.\ \cite{ghp} simplifies to
\be
 D \tau_i = - \rho_{ij} \tau_j.
\label{11e}
\ee
Clearly, the $i = 5$ component is trivial; the $i = 4$ component implies that $\tau_4$ \emph{is independent of} $r$, $\tau_4 = \tau_4(x^{\mu})$.

\subsection{Boost weight 0 components of the Weyl tensor}

In order to continue the integration of the Newman-Penrose equations, we need information about the boost weight 0 components of the Weyl tensor. Their $r$-dependence can be fully determined by the boost weight $+1$ components of the Bianchi identity, eqns.\ B2--B4 of Ref.\ \cite{ghp}. Recall that, in five dimensions, all information regarding the boost weight 0 components of the Weyl tensor is encoded in $\Phi_{ij}$, since
\be
 \Phi_{ijkl} = -2 \left( \Phi^{\mathrm{S}}_{ik} \delta_{jl} - \Phi^{\mathrm{S}}_{il} \delta_{jk} - \Phi^{\mathrm{S}}_{jk} \delta_{il} + \Phi^{\mathrm{S}}_{jl} \delta_{ik} \right) + \Phi \left( \delta_{ik} \delta_{jl} - \delta_{il} \delta_{jk} \right),
\ee
where $\Phi^{\mathrm{S}}_{ij} = \Phi_{(ij)}$.

Consider the Bianchi equation B2 of Ref.\ \cite{ghp},
\be
 D \Phi_{ij} = - \left( \Phi_{ik} + 2 \Phia_{ik} + \Phi \delta_{jk} \right) \rho_{kj}.
\label{A10}
\ee
Integrating each component of this equation with respect to $r$, we find the following results:
\begin{gather}
 \Phi_{44} = f_{44}, \qquad \Phi_{45} = \frac{1}{(r + i \chi)^2} \left[ f_{45} + f_{54} \left( \frac{r^2}{2} + i \chi r \right) \right], \qquad \Phi_{54} = f_{54}, \nonumber \\[2mm]
 \Phi_{55} = \frac{f_{55}}{r + i \chi}, \qquad \Phi_{5\bar{5}} = \frac{1}{(r - i \chi)^3} \left[ f_{5\bar{5}} - f_{44} \left( \frac{r^3}{3} - i \chi r^2 - \chi^2 r \right) \right],
\label{A10:summary}
\end{gather}
where the $f_{ij}$ are complex functions (except for $f_{44}$, which is real), independent of $r$. Note that the $f_{ij}$ are \emph{not} GHP scalars.

Now, we move to eqn.\ B4 of Ref.\ \cite{ghp}, which is purely \emph{algebraic} in our case:
\be
 2 \Phia_{[jk|} \rho_{i|l]} - 2 \Phi_{i[j} \rho_{kl]} + \Phi_{im[jk|} \rho_{m|l]} = 0.
\label{A12}
\ee
Due to the antisymmetry in $jkl$, there are only three real independent components of \eqref{A12}, which are encoded in $ijkl = 445\bar{5}, 545\bar{5}$. The former gives $\chi f_{44} = 0$, whereas the latter implies $f_{54} = 0$. If we now consider eqn.\ B3 of Ref.\ \cite{ghp},
\be
 D \Phi_{ijkl} = 2 \Phi_{[k|i} \rho_{j|l]} - 2 \Phi_{[k|j} \rho_{i|l]} - 4 \Phia_{ij} \rho_{[kl]} - 2 \Phi_{ij[k|m} \rho_{m|l]},
\label{A11}
\ee
we find that all independent components are automatically satisfied, except for $ijkl = 454\bar{5}$ and $ijkl = 5\bar{5}5\bar{5}$, which independently yield $f_{44} = 0$.

In summary, the components of $\Phi_{ij}$ are as follows:\footnote{Some of these results, namely $\Phi_{44} = \Phi_{54} = 0$, have already been obtained in Ref.\ \cite{gspaper}.}
\begin{gather}
 \Phi_{44} = 0, \qquad \Phi_{45} = \frac{f_{45}}{(r + i \chi)^2}, \qquad \Phi_{54} = 0, \nonumber \\[2mm]
 \Phi_{55} = \frac{f_{55}}{r + i \chi}, \qquad \Phi_{5\bar{5}} = \frac{f_{5\bar{5}}}{(r - i \chi)^3}.
\label{Phi:r-dep}
\end{gather}

\subsection{Determining the non-GHP scalars}

The connection components $L_{1i}$, $\stackrel{i}{M}_{j1}$ and $\stackrel{i}{M}_{jk}$ do not transform as GHP scalars. They implicitly enter in the Newman-Penrose equations and Bianchi identity components because they are required for taking GHP derivatives explicitly. Moreover, they also appear in the expressions for the commutators of the basis vectors. In order to determine these objects, we consider the commutators of the GHP derivatives.

Let $V_i$ be an arbitrary GHP scalar with boost weight $b$ and spin weight 1. Consider first the quantity $\left[ \tho, \eth_i \right]V_j$ given by C2 of Ref.\ \cite{ghp}. This gives rise to two different equations. The first comes from the piece that is proportional to the boost weight $b$ of $V_i$ ($b$ is arbitrary, {\textit{cf.}}\ eqn.\ (11b) of Ref.\ \cite{ricci}):
\be
 D L_{1i} = - L_{1j} \rho_{ji}
\label{11b}
\ee
and the second comes from the terms independent of $b$ ({\textit{cf.}}\ eqn.\ (11n) of Ref.\ \cite{ricci}):
\be
 D \stackrel{i}{M}_{jk} \, = - \stackrel{i}{M}_{jl} \rho_{lk}.
\label{11n}
\ee
Integrating eqn.\ \eqref{11b} with respect to $r$, we find
\be
 L_{14} = \lambda_4, \qquad L_{15} = \frac{\lambda_5}{r + i \chi},
\label{L1i:r-dep}
\ee
for $\lambda_i$ independent of $r$ and $\lambda_4$ real. Similarly, integrating eqn.\ \eqref{11n} gives
\begin{equation}
 \stackrel{4}{M}_{54} \, = \, \stackrel{4}{\mu}_{54}, \qquad \stackrel{4}{M}_{55} \, = \frac{\stackrel{4}{\mu}_{55}}{r + i \chi}, \qquad \stackrel{4}{M}_{5\bar{5}} \, = \frac{\stackrel{4}{\mu}_{5\bar{5}}}{r - i \chi},  \qquad
 \stackrel{5}{M}_{\bar{5}4} \, = \, \stackrel{5}{\mu}_{\bar{5}4}, \qquad \stackrel{5}{M}_{\bar{5}5} \, = \frac{\stackrel{5}{\mu}_{\bar{5}5}}{r + i \chi}, \label{11n:summary}
\end{equation}
where the $\stackrel{i}{\mu}_{jk}$ are functions depending on the coordinates $x^{\mu}$ only.

Now, consider the expression $\left[ \tho, \thop \right] V_i$, where $V_i$ is the same GHP scalar as before, with the commutator now given by C1 of Ref.\ \cite{ghp}. As before, this gives rise to two equations, the boost-dependent part ({\textit{cf.}}\ eqn.\ (11a) of Ref.\ \cite{ricci})
\be
 D L_{11} = - L_{14} \tau_4 - \Phi + \frac{\Lambda}{4} \label{11a}
\ee
and the boost-independent part ({\textit{cf.}}\ eqn.\ (11m) of Ref.\ \cite{ricci})
\be
 D \stackrel{i}{M}_{j1} = - \stackrel{i}{M}_{j4} \tau_4 - 2 \Phia_{ij}.
\label{11m}
\ee
Using our knowledge of $\Phi_{ij}$ from eqn.\ \eqref{Phi:r-dep}, we can then fully integrate these equations to obtain
\be
 L_{11} = \lambda_1 + \left( \frac{\Lambda}{4} - \lambda_4 \tau_4 \right) r + \frac{f_{5\bar{5}}}{2(r - i \chi)^2} + \frac{f_{\bar{5}5}}{2(r + i \chi)^2}
\label{L11:r-dep}
\ee
and
\bea
 \stackrel{4}{M}_{51} &=& \stackrel{4}{\mu}_{51} - \tau_4 \stackrel{4}{\mu}_{54} (r + i \chi) + \frac{f_{45}}{r + i \chi}, \nonumber \\
 \stackrel{5}{M}_{\bar{5}1} &=& \stackrel{5}{\mu}_{\bar{5}1} - \tau_4 \stackrel{5}{\mu}_{\bar{5}4} r + \frac{f_{5\bar{5}}}{2(r - i \chi)^2} - \frac{f_{\bar{5}5}}{2(r + i \chi)^2},
\label{11m:summary}
\eea
respectively. Here real $\lambda_1$ and complex $\stackrel{i}{\mu}_{j1}$ are independent of $r$. This fully determines the $r$-dependence of the non-GHP scalars.

\subsection{Optical matrix $\rho'_{ij}$ of $n$}

We return to our analysis of the NP equations and consider eqn.\ NP4$'$ of Ref.\ \cite{ghp},
\be
 D \rho'_{ij} = - \rho'_{ik} \rho_{kj} - \Phi_{ji} - \frac{\Lambda}{4} \delta_{ij},
\label{11j}
\ee
which determines the $r$-dependence of $\rho'_{ij}$. Since we know the full $r$-dependence of $\Phi_{ij}$ from eqn.\ \eqref{Phi:r-dep}, we can now integrate each component of this equation to obtain:
\begin{gather}
 \rho'_{44} = A_{44} - \frac{\Lambda}{4}r, \qquad \rho'_{45} = \frac{A_{45}}{r + i \chi}, \qquad \rho'_{54} = A_{54} + \frac{f_{45}}{r + i \chi}, \nonumber \\[2mm]
 \rho'_{55} = \frac{A_{55}}{r + i \chi} - f_{55}, \qquad \rho'_{5\bar{5}} = \frac{1}{r - i \chi} \left[ A_{5\bar{5}} + \frac{f_{\bar{5}5}\, r}{(r + i \chi)^2} - \frac{\Lambda}{8}(r + i \chi)(r - 3 i \chi) \right], \label{rhoprime:r-dep}
\end{gather}
where the $A_{ij}$ are complex functions (except for $A_{44}$, which is real) that do not depend on $r$.

\subsection{The $r$-dependence of the basis vectors}

Let us now consider the commutator of $\ell$ with the other basis vectors. The reason for doing this is that, in our local coordinates, this will determine the $r$-derivative of each component of $n$ and the $m_i$, which can then be used to determine the $r$-dependence of the entire basis. Consider first the commutators
\be
 \left[ \ell, m_i \right] = - \left( \tau'_i + L_{1i} \right) \ell + ( \stackrel{i}{M}_{j0} - \rho_{ji} )\, m_j,
\label{l:mi:commutator}
\ee
where the right hand side is determined by projecting the covariant expression of the commutator (in the Levi-Civita connection) along the basis and using the definition of the various connection components. In our choice of coordinates and basis, eqn.\ \eqref{l:mi:commutator} reduces to
\be
 D m_i^A = - L_{1i}\, \ell^A - \rho_{ji}\, m_j^A,
\label{l:mi:commutator:coordinates}
\ee
where we are using an uppercase Latin index to denote collectively all the coordinate indices corresponding to $r, x^{\mu}$. Let us first look at $i = 4$ component,
\be
 D m_4^A = - \lambda_4\; \ell^A.
\ee
The $r$-component is
\be
 D m_4^r = - \lambda_4,
\ee
with solution
\be
 m_4^r = E_4 - \lambda_4\, r,
\ee
where real $E_4$ does not depend on $r$. The $\mu$-components give simply
\be
 m_4^{\mu} = (m_4^0)^{\mu} (x).
\ee
Consider a boost $\ell \rightarrow \beta \ell, n \rightarrow \beta^{-1} n$. Previously, we used a boost transformation to rescale $\ell$ such that it is tangent to affinely parameterised geodesics, i.e.\ $L_{10} = 0$. This leaves a residual freedom of performing a boost with $D \beta = 0$. Under such a transformation, $\tau_4$ remains invariant, while $\lambda_4 \rightarrow \lambda_4 + m_4^0(\beta)$. Hence, we can use this transformation to set 
\be
 \lambda_4 = \tau_4,
\ee
which will be assumed from now on. Thus, 
\be
 m_4 = \left( E_4 - \tau_4\, r \right) \frac{\partial}{\partial r} + (m_4^0)^{\mu} \frac{\partial}{\partial x^{\mu}}.
\label{m4:r-dep}
\ee

Now we go back to eqn.\ \eqref{l:mi:commutator:coordinates} and take $i = 5$:
\be
 D m_5^A = - \frac{1}{r + i \chi}(m_5^A + \lambda_5\, \ell^A).
\ee
For $A = r$ the second term introduces an inhomogeneous contribution, and the solution is
\be
 m_5^r = \frac{E_5}{r + i \chi} - \lambda_5,
\ee
where complex $E_5$ does not depend on $r$. The $A = \mu$ components are homogeneous equations solved simply by
\be
 m_5^{\mu} = \frac{(m_5^0)^{\mu}}{r + i \chi},
\ee
where complex $(m_5^0)^{\mu}$ is independent of $r$. Thus,
\be
 m_5 = \left( \frac{E_5}{r + i \chi} - \lambda_5 \right) \frac{\partial}{\partial r} + \frac{(m_5^0)^{\mu}}{r + i \chi} \frac{\partial}{\partial x^{\mu}}.
\label{m5:r-dep}
\ee

Finally, we consider the commutator
\be
 \left[ \ell, n \right] = - L_{11} \ell - L_{10}\, n - \left( \tau_i - \tau'_i \right) m_i.
\label{l:n:commutator}
\ee
In our choice of coordinates and basis, this is simply
\be
 D n^A = - L_{11} \ell^A - \tau_4 m_4^A.
\label{l:n:commutator:coordinates}
\ee
Integrating gives
\be \label{nr}
 n^r = E_1 - \left( \lambda_1 + \tau_4 E_4 \right) r - \left( \frac{\Lambda}{8} - \tau_4^2 \right) (r^2 + \chi^2) + \frac{f_{5\bar{5}}}{2(r - i \chi)} + \frac{f_{\bar{5}5}}{2(r + i \chi)},
\ee
where real $E_1$ does not depend on $r$ and
\be
 n^{\mu} = (n^0)^{\mu} - r\, \tau_4\, (m_4^0)^{\mu},
\ee
where real $(n^0)^{\mu}$ is independent of $r$. In summary, for $n$ we have
\bea
\label{nsol}
 n &=& \left[ E_1 - \left( \lambda_1 + \tau_4 E_4 \right) r - \left( \frac{\Lambda}{8} - \tau_4^2 \right) (r^2 + \chi^2) + \frac{f_{5\bar{5}}}{2(r - i \chi)} + \frac{f_{\bar{5}5}}{2(r + i \chi)} \right] \frac{\partial}{\partial r} \nonumber \\
   &\quad& + \left[ (n^0)^{\mu} - r\, \tau_4\, (m_4^0)^{\mu} \right] \frac{\partial}{\partial x^{\mu}}, \label{n:r-dep}
\eea
thereby completing the determination of the $r$-dependence of the basis vectors. 

%%%%%%%%%%%%%%%%%%%%%%%%%%%%%%%%%%%%%%%

\section{Differential and Algebraic Constraints} \label{sec:constraint}

%%%%%%%%%%%%%%%%%%%%%%%%%%%%%%%%%%%%%%%%

At the end of the previous section, we determined the full $r$-dependence of the basis vectors. This also determines the $r$-dependence of the metric via the relation
\be
 g^{ab} = 2 \, \ell^{(a}\,  n^{b)} + m_i^{a}\,  m_i^b.
\ee
We now need to examine how the Einstein equation constrains the dependence on the other coordinates $x^\mu$. There are several equations that we have not considered yet, and these provide additional information on the curvature and hence contain other components of the Einstein equation. 

\subsection{Equation NP3}

Since the $m_i$ are simpler than $n$, and since $\rho_{ij}$ is the simplest object we have, it is easiest to start with an equation involving $\eth_i \rho_{jk}$. Such an equation is NP3 of Ref.\ \cite{ghp}, which in a more explicit form is
\begin{equation}
 \delta_j \rho_{ik} - \delta_k \rho_{ij} \,=\, L_{1j} \rho_{ik} - L_{1k} \rho_{ij} + \tau_i ( \rho_{jk} - \rho_{kj} )  + \rho_{il} ( \stackrel{l}{M}_{jk} - \stackrel{l}{M}_{kj} ) + \rho_{lj} \stackrel{l}{M}_{ik} - \rho_{lk} \stackrel{l}{M}_{ij}.				\label{11k}
\end{equation}
Because of the antisymmetry in $jk$, there are, in total, five components to consider. We start with $ijk = 445$, which gives $\rho_{\bar{5}5} \stackrel{4}{M}_{54} = 0$. Hence, from eqn.\ \eqref{11n:summary}
\be
\stackrel{4}{\mu}_{54} \, = 0.
\ee
Next we take $ijk = 45\bar{5}$, giving
\be
 \stackrel{4}{\mu}_{5\bar{5}} - \stackrel{4}{\mu}_{\bar{5}5} \, = 2 i \chi \tau_4.
\label{11k:455bar}
\ee
For $ijk = 545$, we get $- \rho_{5\bar{5}} \stackrel{4}{M}_{55} = 0$, hence, again from eqn.\ \eqref{11n:summary},
\be
\stackrel{4}{\mu}_{55} = 0.
\ee
The next component we consider is $ijk = 54\bar{5}$, which gives
\be
 - E_4 + i m_4^0(\chi) = - i \chi \tau_4 - \stackrel{4}{\mu}_{5\bar{5}}.
\label{11k:545bar}
\ee
Hence,
\be
 \stackrel{4}{\mu}_{5\bar{5}} \, = -i \chi \tau_4 + E_4 - i m_4^0(\chi).
\ee
Substituting this into eqn.\ \eqref{11k:455bar} gives
\be
 m_4^0(\chi) = - 2 \tau_4 \chi.
\ee
Substituting this back into \eqref{11k:545bar}, we then find that
\be
 \stackrel{4}{\mu}_{5\bar{5}} \, = E_4 + i \chi \tau_4.
\ee
Finally, the last component $ijk = 55\bar{5}$ reduces to
\be
 m_5^0(\chi) = -i E_5 - 2 \chi \lambda_5.
\ee
A summary of the results obtained from eqn.\ \eqref{11k} is as follows:
\begin{equation}
 \stackrel{4}{\mu}_{54} \, = 0, \qquad \stackrel{4}{\mu}_{55}\, = 0, \qquad \stackrel{4}{\mu}_{5\bar{5}} = E_4 + i \chi \tau_4, \qquad 
 m_4^0(\chi) = - 2 \tau_4 \chi, \qquad m_5^0(\chi) = -i E_5 - 2 \chi \lambda_5.
\label{11k:summary}
\end{equation}

\subsection{Equation NP4}

Now, we move to eqn.\ NP4 of Ref.\ \cite{ghp}:
\be
 \Delta \rho_{ij} - \delta_j \tau_i = L_{11} \rho_{ij} - \tau_i \tau_j + \tau_4 \stackrel{4}{M}_{ij} - \rho_{kj} \stackrel{k}{M}_{i1} - \rho_{ik} ( \rho'_{kj} + \stackrel{k}{M}_{j1} ) - \Phi_{ij} - \frac{\Lambda}{4} \delta_{ij}.
\label{11i}
\ee
The $ij = 44$ component gives
\be
 m_4^0(\tau_4) = \tau_4^2 + \frac{\Lambda}{4},
\ee
while $ij = 45$ gives
\be
 m_5^0(\tau_4) = - \stackrel{4}{\mu}_{51}.
\ee
Taking $ij = 54$, we obtain the relation
\be
 \rho'_{54} \, - \stackrel{4}{M}_{51} \, = 0,
\ee
which, using eqns.\ \eqref{11m:summary} and \eqref{rhoprime:r-dep}, reduces to
\be
 \stackrel{4}{\mu}_{51} \, = A_{54}.
\ee
The $ij = 55$ component gives
\be
 A_{55} = 2 i \chi f_{55}.
\ee
The only remaining component left to consider is $ij = 5\bar{5}$, which gives a more complicated relation:
\be
i n^0(\chi) + i \chi (\lambda_1 + \tau_4 E_4)  + A_{5\bar{5}} - E_1 - 2 \left(\tau_4^2 + \frac{\Lambda}{4} \right) \chi^2 = 0.
\ee
By taking the imaginary part, we obtain an expression for $n^0(\chi)$:
\be
 n^0(\chi) = \frac{i \left( A_{5\bar{5}} - A_{\bar{5}5} \right)}{2} - \chi \left( \lambda_1 + \tau_4 E_4 \right).
\ee
The real part, on the other hand, gives an algebraic relation
\be
A_{5\bar{5}} + A_{\bar{5}5} - 2 E_1 - ( \Lambda + 4 \tau_4^2 ) \chi^2 = 0.
\ee
In summary, the information obtained from NP4, eqn.\ \eqref{11i}, is as follows:
\begin{gather}
 m_4^0(\tau_4) = \tau_4^2 + \frac{\Lambda}{4}, \qquad m_5^0(\tau_4) = - A_{54}, \qquad \stackrel{4}{\mu}_{51} \, = A_{54}, \qquad A_{55} = 2 i \chi f_{55}, \nonumber \\[2mm]
  n^0(\chi) = \frac{i \left( A_{5\bar{5}} - A_{\bar{5}5} \right)}{2} - \chi \left( \lambda_1 + \tau_4 E_4 \right), \qquad  A_{5\bar{5}} + A_{\bar{5}5} - 2 E_1 - ( \Lambda + 4 \tau_4^2 ) \chi^2 = 0.
\label{11i:summary}
\end{gather}

\subsection{Commutator C3}

We have exhausted the Newman-Penrose equations that involve only quantities that we already know. In order to proceed with the remaining ones, we need information about the GHP scalars with negative boost weight such as $\kappa'_i$ and $\Psi'_{ijk}$. However, we can investigate other equations that do not involve such unknown objects. One such equation comes from applying the commutator $\left[ \eth_i, \eth_j \right]$ to some arbitrary GHP scalar $V_i$ of spin weight 1 but arbitrary boost weight, as before. This is determined by eqn.\ C3 of Ref.\ \cite{ghp}. Once again, this gives us two different equations. The first comes from the piece that depends on the boost weight $b$ of $V_i$ ({\textit{cf.}}\ eqn.\ (11d) of Ref.\ \cite{ricci}),
\be
 \delta_j L_{1i} - \delta_i L_{1j} = - L_{11} \left( \rho_{ij} - \rho_{ji} \right) - L_{1k}( \stackrel{k}{M}_{ij} - \stackrel{k}{M}_{ji}) - \rho_{kj} \rho'_{ki} + \rho_{ki} \rho'_{kj} + \Phi_{ij} - \Phi_{ji}.
\label{11d}
\ee
The antisymmetry in $ij$ implies that there are only two independent components to consider: $45$ and $5\bar{5}$. For $ij = 45$, we find
\be
 m_4^0(\lambda_5) = (\stackrel{5}{\mu}_{\bar{5}4} - \tau_4 ) \lambda_5.
\label{m40:lambda5}
\ee
For the other component, $ij = 5\bar{5}$, we get
\be
 \mbar_5^0(\lambda_5) - m_5^0(\bar{\lambda}_5) = - 2 i \chi \lambda_1 - 2 i \chi \tau_4 E_4 - \lambda_5 \stackrel{\bar{5}}{\mu}_{5\bar{5}} + \bar{\lambda}_5 \stackrel{5}{\mu}_{\bar{5}5} + A_{5\bar{5}} - A_{\bar{5}5}.
\ee

On the other hand, the piece of C3 that is independent of the boost weight is ({\textit{cf.}}\ eqn.\ (11p) of Ref.\ \cite{ricci})
\bea
 \delta_k \stackrel{i}{M}_{jl} - \delta_l \stackrel{i}{M}_{jk} &=& \rho'_{il} \rho_{jk} - \rho'_{ik} \rho_{jl} + \rho_{il} \rho'_{jk} - \rho_{ik} \rho'_{jl} + \left( \rho_{kl} - \rho_{lk} \right) \stackrel{i}{M}_{j1} + \stackrel{i}{M}_{pk} \stackrel{p}{M}_{jl} - \stackrel{i}{M}_{pl} \stackrel{p}{M}_{jk} \nonumber \\
 &\quad& + \stackrel{i}{M}_{jp} ( \stackrel{p}{M}_{kl} - \stackrel{p}{M}_{lk} ) - \Phi_{ijkl} - \frac{\Lambda}{4} \left( \delta_{ik} \delta_{jl} - \delta_{il} \delta_{jk} \right).
\label{11p}
\eea
Due to antisymmetry in both $ij$ and $kl$, the independent components are $ijkl = 4545$, $454\bar{5}$, $455\bar{5}$, $5\bar{5}45$ and $5\bar{5}5\bar{5}$. The $ijkl = 4545$ component simply gives
\be
 f_{55} = 0.
\ee
The $454\bar{5}$ component gives
\be
 m_4^0(E_4) = - (A_{44} + \tau_4 E_4).
\ee
For $ijkl = 455\bar{5}$, we get
\be
 m_5^0(E_4) = - A_{45} + f_{45} - \lambda_5 E_4 - \tau_4 E_5 + i \chi \left( 3 A_{54} + \lambda_5 \tau_4 \right).
\ee
Taking now $ijkl = 5\bar{5}45$, we find
\be
 m_4^0 ( \stackrel{5}{\mu}_{\bar{5}5} ) - m_5^0 ( \stackrel{5}{\mu}_{\bar{5}4} ) + A_{54} + (\tau_4-\stackrel{5}{\mu}_{\bar{5}4}) \stackrel{5}{\mu}_{\bar{5}5} = 0.
\ee
Finally, $ijkl = 5\bar{5}5\bar{5}$ gives
\begin{align}
m_5^0 ( \stackrel{\bar{5}}{\mu}_{5\bar{5}} ) + \mbar_5^0 ( \stackrel{5}{\mu}_{\bar{5}5} ) + A_{5\bar{5}} + A_{\bar{5}5} + E_4^2 + \lambda_5 \stackrel{\bar{5}}{\mu}_{5\bar{5}}& + \bar{\lambda}_5 \stackrel{5}{\mu}_{\bar{5}5} \nonumber \\[2mm]
    + 2 \stackrel{5}{\mu}_{\bar{5}5} \stackrel{\bar{5}}{\mu}_{5\bar{5}} &+ 2 i \chi ( \stackrel{5}{\mu}_{\bar{5}1} + E_4 \stackrel{5}{\mu}_{\bar{5}4} ) - \frac{\Lambda \chi^2}{2} + \chi^2 \tau_4^2 \,=\, 0.
\end{align}
In summary, we have:
\bea
 f_{55} &=& 0, \nonumber \\[1mm]
 m_4^0(E_4) &=& - (A_{44} + \tau_4 E_4), \nonumber \\[1mm]
 m_5^0(E_4) &=& - A_{45} + f_{45} - \lambda_5 E_4 - \tau_4 E_5 + i \chi \left( 3 A_{54} + \lambda_5 \tau_4 \right), \nonumber \\[1mm]
 m_4^0(\lambda_5) &=&  (\stackrel{5}{\mu}_{\bar{5}4} - \tau_4 ) \lambda_5, \nonumber \\[1mm]
 m_5^0(\bar{\lambda}_5) - \mbar_5^0(\lambda_5) &=& 2 i \chi \lambda_1 + 2 i \chi \tau_4 E_4 + \lambda_5 \stackrel{\bar{5}}{\mu}_{5\bar{5}} - \bar{\lambda}_5 \stackrel{5}{\mu}_{\bar{5}5} - A_{5\bar{5}} + A_{\bar{5}5}, \nonumber \\[1mm]
 m_4^0 ( \stackrel{5}{\mu}_{\bar{5}5} ) - m_5^0 ( \stackrel{5}{\mu}_{\bar{5}4} ) &=&  - A_{54} +  (\stackrel{5}{\mu}_{\bar{5}4} - \tau_4) \stackrel{5}{\mu}_{\bar{5}5}, \nonumber \\[1mm]
 m_5^0 ( \stackrel{\bar{5}}{\mu}_{5\bar{5}} ) + \mbar_5^0 ( \stackrel{5}{\mu}_{\bar{5}5} ) &=&  - A_{5\bar{5}} - A_{\bar{5}5} - E_4^2 - \lambda_5 \stackrel{\bar{5}}{\mu}_{5\bar{5}} - \bar{\lambda}_5 \stackrel{5}{\mu}_{\bar{5}5} \nonumber \\[2mm]
   &\quad& - 2 \stackrel{5}{\mu}_{\bar{5}5} \stackrel{\bar{5}}{\mu}_{5\bar{5}} - 2 i \chi ( \stackrel{5}{\mu}_{\bar{5}1} + E_4 \stackrel{5}{\mu}_{\bar{5}4} ) + \left( \frac{\Lambda}{2} - \tau_4^2 \right) \chi^2 . \label{11p:summary}
\eea

\subsection{Commutators $\left[ m_i, m_j \right]$}

We will now consider the commutators
\be
 \left[ m_i, m_j \right] = 2 \rho'_{[ij]}\, \ell + 2 \rho_{[ij]}\, n + 2 \stackrel{k}{M}_{[ij]}\, m_k.
\label{mi:mj:commutator}
\ee
Taking $ij = 45$ and using the results above gives
\be
 \left[ m_4, m_5 \right] = ( \rho'_{45} - \rho'_{54} )\, \ell - ( \stackrel{4}{M}_{\bar{5}5} - \stackrel{5}{M}_{\bar{5}4} )\, m_5.
\label{m4:m5:commutator}
\ee
From the $r$-component, we get
\be
m_4^0(E_5) + (3 \tau_4 - \stackrel{5}{\mu}_{\bar{5}4}) E_5  - 2 i \chi A_{54} = 0,
\ee
while the $\mu$-components give
\be
 \left[ m_4^0, m_5^0 \right] = ( \stackrel{5}{\mu}_{\bar{5}4} - \tau_4 )\, m_5^0.
\ee

Setting $ij = 5\bar{5}$ in eqn.\ \eqref{mi:mj:commutator} gives:
\be
 \left[ m_5, \mbar_5 \right] = 2 \rho'_{[5\bar{5}]}\, \ell + 2 \rho_{[5\bar{5}]}\, n + 2 \stackrel{k}{M}_{[5\bar{5}]} m_k.
\label{m5:m5bar:commutator}
\ee
The $r$-component of the above equation gives
\begin{align}
m_5^0(\bar{E}_5) - \mbar_5^0(E_5) &+ i \chi \left[m_5^0(\bar{\lambda}_5) + \mbar_5^0(\lambda_5)\right] + \lambda_5 \bar{E}_5 - \bar{\lambda}_5 E_5 + f_{5\bar{5}} - f_{\bar{5}5} \nonumber \\
   & -2 i \chi (2 E_1 + E_4^2 +3 \chi^2 \tau_4^2) - E_5 \stackrel{\bar{5}}{\mu}_{5\bar{5}} + \bar{E}_5 \stackrel{5}{\mu}_{\bar{5}5} + i \chi \lambda_5 \stackrel{\bar{5}}{\mu}_{5\bar{5}} + i \chi \bar{\lambda}_5 \stackrel{5}{\mu}_{\bar{5}5} = 0,
\end{align}
where we have used \eqref{11i:summary}.  On the other hand, the $\mu$-components give the commutator
\be
 \left[ m_5^0, \mbar_5^0 \right] = 2 i \chi\, n^0 + 2 i \chi E_4\, m_4^0 + ( \bar{\lambda}_5 + \stackrel{\bar{5}}{\mu}_{5\bar{5}} )\, m_5^0 -  ( \lambda_5 + \stackrel{5}{\mu}_{\bar{5}5} )\, \mbar_5^0.
\ee

In summary, from the commutators \eqref{mi:mj:commutator}, we obtain the following relations:
\bea
 m_4^0(E_5) &=& 2 i \chi A_{54} - (3 \tau_4 - \stackrel{5}{\mu}_{\bar{5}4}) E_5, \nonumber \\[1mm]
 \mbar_5^0(E_5) - m_5^0(\bar{E}_5) - i \chi \left[m_5^0(\bar{\lambda}_5) + \mbar_5^0(\lambda_5)\right] &=& \lambda_5 \bar{E}_5 - \bar{\lambda}_5 E_5 + f_{5\bar{5}} - f_{\bar{5}5} - E_5 \stackrel{\bar{5}}{\mu}_{5\bar{5}} + \bar{E}_5 \stackrel{5}{\mu}_{\bar{5}5} \nonumber \\
   &\quad&  + i \chi (\lambda_5 \stackrel{\bar{5}}{\mu}_{5\bar{5}} + \bar{\lambda}_5 \stackrel{5}{\mu}_{\bar{5}5} - 4 E_1 -2 E_4^2 -6 \chi^2 \tau_4^2)\nonumber \\
 {} &\quad& {} \label{mi:mj:commutator:summary:a} 
\eea
and the commutators
\bea
 \left[ m_4^0, m_5^0 \right] &=& ( \stackrel{5}{\mu}_{\bar{5}4} - \tau_4 )\, m_5^0, \nonumber \\[1mm]
 \left[ m_5^0, \mbar_5^0 \right] &=& 2 i \chi\, n^0 + 2 i \chi E_4\, m_4^0 + ( \bar{\lambda}_5 + \stackrel{\bar{5}}{\mu}_{5\bar{5}} )\, m_5^0 -  ( \lambda_5 + \stackrel{5}{\mu}_{\bar{5}5} )\, \mbar_5^0. \label{mi:mj:commutator:summary:b}
\eea

\subsection{Calculation of $\Psi'_{ijk}$ from the Bianchi identities}

In order to proceed with analysing the remaining Newman-Penrose equations, we need information regarding the boost weight $-1$ components of the Weyl tensor.  The boost weight 0 components of the Bianchi identity, eqns.\ B5--B7 and B5$'$ of Ref.\ \cite{ghp} can be used to calculate $\Psi'_{ijk}$. From the symmetries of $\Psi'_{ijk}$, $\Psi'_{ijk} = - \Psi'_{ikj}$ and $\Psi'_{[ijk]} = 0$, we notice that the independent components are
\be
 \Psi'_{445}, \qquad \Psi'_{545}, \qquad \Psi'_{54\bar{5}}, \qquad \Psi'_{55\bar{5}}.
\ee

We start with eqn.\ B5 of Ref.\ \cite{ghp},
\be
 - 2 \eth_{[j|} \Phi_{i|k]} = \left( 2 \Phi_{i[j} \delta_{k]l} - 2 \delta_{il} \Phia_{jk} - \Phi_{iljk} \right) \tau_l + 2 \left( \Psi'_{[j|} \delta_{il} - \Psi'_{[j|il} \right) \rho_{l|k]}.
\label{A13}
\ee
Due to antisymmetry in $jk$, the independent components that we need to consider are $445$, $45\bar{5}$, $545$, $54\bar{5}$ and $55\bar{5}$. We start with $ijk = 445$, which gives
\be
 \Psi'_{445} = \frac{m_4^0(f_{45})}{r + i \chi} - \frac{f_{45}}{(r + i \chi)^2} \left[ E_4  - i \chi \tau_4 + \stackrel{5}{\mu}_{\bar{5}4}(r + i \chi) \right].
\label{A13:445}
\ee
The $ijk = 545$ component is automatically satisfied because $f_{54} =f_{55} = 0$.
Setting $ijk = 54\bar{5}$, we find
\be
 \Psi'_{54\bar{5}} = - \frac{m_4^0(f_{\bar{5}5})}{2 (r + i \chi)^2} + \frac{f_{\bar{5}5}}{(r + i \chi)^3} \left[ E_4 - \tau_4 (r +2 i \chi) \right].
\label{A13:545bar}
\ee
Substituting the above result into the $ijk = 45\bar{5}$ component, gives
\begin{equation}
 \frac{m_4^0(f_{5\bar{5}}) + 4 \tau_4 f_{5\bar{5}} - 2 m_5^0(f_{4\bar{5}}) - 2 ( 2\lambda_5 + \stackrel{5}{\mu}_{\bar{5}5}) f_{4\bar{5}}}{2 (r - i \chi)^2 (r + i \chi)}  - \text{c.c.} \,=\, 0. \label{A13:455bar}
\end{equation}
From the $r$-dependence of the above equation, we conclude that\footnote{If $\chi = 0$, then the $r$-dependence in \eqref{A13:455bar} can be factored out and we can only conclude that the left hand side of \eqref{A14:summary} is real. However, comparison with e.g.\ the $ijklm = 5\bar{5}45\bar{5}$ component of eqn.\ B6 of Ref.\ \cite{ghp} implies that eqn.\ \eqref{A14:summary} is true also when $\chi = 0$.}
\be
m_4^0(f_{5\bar{5}}) + 4 \tau_4 f_{5\bar{5}} - 2 m_5^0(f_{4\bar{5}}) - 2 ( 2\lambda_5 + \stackrel{5}{\mu}_{\bar{5}5}) f_{4\bar{5}} \,=\, 0.
\label{A14:summary}
\ee
Finally, we consider the $ijk = 55\bar{5}$ component and use the result \eqref{A13:445} for $\Psi'_{445}$, which gives
\be
 (r - i \chi)^2 \left[ m_4^0(f_{45}) +  ( \tau_4 - \stackrel{5}{\mu}_{\bar{5}4} ) f_{45} \right] = m_5^0(f_{5\bar{5}}) + 3 \lambda_5 f_{5\bar{5}}.
\label{A13:555bar}
\ee
Analysing the $r$-dependence, we immediately see that both the left hand side and right hand side must vanish independently. We therefore have
\be
 m_4^0(f_{45}) = (\stackrel{5}{\mu}_{\bar{5}4} - \tau_4) f_{45}
\label{m40:f45}
\ee
and
\be
  m_5^0(f_{5\bar{5}}) = - 3 \lambda_5 f_{5\bar{5}}.
\ee
Note that, using eqn.\ \eqref{m40:f45}, we can rewrite $\Psi'_{445}$ in a simpler form:
\be
\Psi'_{445} = - \frac{(E_4 + \tau_4 r)}{(r + i \chi)^2} f_{45}.
\ee
This exhausts eqn.\ \eqref{A13}. The summary is as follows:
\begin{gather}
 \Psi'_{445} \,=\, - \frac{(E_4 + \tau_4 r)}{(r + i \chi)^2} f_{45}, \qquad
 \Psi'_{54\bar{5}} \,=\, - \frac{m_4^0(f_{\bar{5}5})}{2 (r + i \chi)^2} 
 + \frac{f_{\bar{5}5}}{ (r + i \chi)^3} \left[  E_4 - \tau_4 (r +2 i \chi) \right], \notag \\[1mm]
 m_5^0(f_{5\bar{5}}) \,=\, - 3 \lambda_5 f_{5\bar{5}}, \qquad
 m_4^0(f_{45}) \,=\, (\stackrel{5}{\mu}_{\bar{5}4} - \tau_4) f_{45}, \notag \\[2mm]
2 m_5^0(f_{4\bar{5}}) - m_4^0(f_{5\bar{5}}) \,=\,    - 2 f_{4\bar{5}} ( 2\lambda_5 + \stackrel{5}{\mu}_{\bar{5}5} ) + 4 \tau_4 f_{5\bar{5}}.
 \label{A13:summary}
\end{gather}

Using the results from B5 above, one can verify that eqns.\ B6 and B7 of Ref.\ \cite{ghp} are automatically satisfied. The only remaining boost weight 0 component of the Bianchi equation is then B5$'$:
\be
 D \Psi'_{ijk} -2 \eth_{[j} \Phi_{k]i} = 2 \left( \Psi'_i \delta_{[j|l} - \Psi'_{i[j|l} \right) \rho_{l|k]}.
\label{A16}
\ee
Once again, antisymmetry in $jk$ implies that the independent components are $ijk = 445$, $45\bar{5}$, $545$, $54\bar{5}$, $55\bar{5}$. Of these, $445$, $45\bar{5} $ and $54\bar{5}$ are automatically satisfied using the results from B5 above. On the other hand, we can integrate $ijk = 545$ and $ijk = 55\bar{5}$ with respect to $r$ to find the remaining components of $\Psi'_{ijk}$, which we simplify using the equation for $m_4^0(f_{45})$, \eqref{m40:f45}. The former gives
\be
 \Psi'_{545} = \frac{B_{545}}{r + i \chi} + \frac{m_5^0(f_{45}) + (2 \lambda_5 - \stackrel{5}{\mu}_{\bar{5}5}) f_{45}}{(r + i \chi)^2} - \frac{2 (E_5 - i \chi \lambda_5) f_{45}}{(r + i \chi)^3},
\ee
where $B_{545}$ is some complex function independent of $r$, while the latter integrates to
\be
 \Psi'_{55\bar{5}} = \frac{B_{55\bar{5}}}{(r + i \chi)^2} - \frac{m_5^0(f_{\bar{5}5}) + 3 \lambda_5 f_{\bar{5}5}}{(r + i \chi)^3} + \frac{3  (E_5 - i \chi \lambda_5) f_{\bar{5}5}}{(r + i \chi)^4},
\ee
where $B_{55\bar{5}}$ is another $r$-independent, complex, function.

\subsection{Newman-Penrose equations with boost weight $-1$}

We are now in a position to look at the other Newman-Penrose equations. We start with NP2$'$ of Ref.\ \cite{ghp},
\be
 D \kappa'_i = - \rho'_{ij} \tau_j + \Psi'_i.
\label{11f}
\ee
Integrating the $i = 4$ component, we find
\bea
 \kappa'_4 &=& G_4 - \tau_4 A_{44} r + \frac{\Lambda}{8} \tau_4 (r^2 + \chi^2) 
 + \frac{m_4^0(f_{5\bar{5}}) + 2 \tau_4 f_{5\bar{5}}}{2(r - i \chi)} + \frac{m_4^0(f_{\bar{5}5}) + 2 \tau_4 f_{\bar{5}5}}{2(r + i \chi)} \nonumber \\[1mm]
           &\quad& - \frac{ (E_4 + i \chi \tau_4)f_{5\bar{5}}}{2 (r - i \chi)^2} - \frac{ (E_4 - i \chi \tau_4)f_{\bar{5}5}}{2 (r + i \chi)^2},
\eea
where $G_4$ is real and does not depend on $r$. Similarly, the $i = 5$ component integrates to
\be
 \kappa'_5 = G_5 - \tau_4 A_{54} (r + i \chi) - \frac{B_{55\bar{5}} + (E_4 - i \chi \tau_4) f_{45}}{r + i \chi} + \frac{m_5^0(f_{\bar{5}5}) + 3 \lambda_5 f_{\bar{5}5}}{2 (r + i \chi)^2} -  \frac{(E_5 - i \chi \lambda_5)f_{\bar{5}5}}{(r + i \chi)^3},
\ee
where complex $G_5$ depends only on $x^{\mu}$.

The next equation we look at is eqn.\ NP3$'$ of Ref.\ \cite{ghp},
\be
 \delta_j \rho'_{ik} - \delta_k \rho'_{ij} = - L_{1j} \rho'_{ik} + L_{1k} \rho'_{ij} + \kappa'_i \left( \rho_{jk} - \rho_{kj} \right) + \rho'_{il} ( \stackrel{l}{M}_{jk} - \stackrel{l}{M}_{kj} ) + \rho'_{lj} \stackrel{l}{M}_{ik} - \rho'_{lk} \stackrel{l}{M}_{ij} - \Psi'_{ijk}.
\label{11l}
\ee
Because of the antisymmetry in $jk$, the only independent components are $ijk = 445$, $45\bar{5}$, $545$, $54\bar{5}$ and $55\bar{5}$. Let us start with $ijk = 445$, which gives
\be
m_4^0(A_{45}) - m_5^0(A_{44}) - f_{45} \tau_4 - A_{44} \lambda_5 + A_{45} ( 2 \tau_4 - \stackrel{5}{\mu}_{\bar{5}4} ) + \frac{\Lambda}{4} (E_5 - i \chi \lambda_5) + A_{54} (E_4 - i \chi \tau_4) \,=\, 0.
\ee
Now taking $ijk = 45\bar{5}$ gives
\begin{align}
 &2 (A_{5\bar{5}} - A_{\bar{5}5}) E_4 - m_4^0(f_{5\bar{5}}) + m_4^0(f_{\bar{5}5}) - 2 m_5^0 (A_{4\bar{5}}) + 2 \mbar_5^0(A_{45}) - 2 A_{4\bar{5}} ( 2 \lambda_5 + \stackrel{5}{\mu}_{\bar{5}5} ) \nonumber \\[1mm]
  &  + 2 A_{45} ( 2 \bar{\lambda}_5 + \stackrel{\bar{5}}{\mu}_{5\bar{5}} ) - 2 \tau_4 (f_{5\bar{5}} - f_{\bar{5}5}) + 2 i \chi \left[ 2 A_{44} E_4 + 2 G_4 - (A_{5\bar{5}} + A_{\bar{5}5} - \Lambda \chi^2) \tau_4 \right] = 0. 
\end{align}
For $ijk = 545$ we get a very simple equation
\be
 B_{545} = m_5^0(A_{54}) + A_{54} ( \lambda_5 - \stackrel{5}{\mu}_{\bar{5}5} ).
\ee
Next we consider $ijk = 54\bar{5}$, which results in
\be
\mbar_5^0(A_{54}) - m_4^0(A_{5\bar{5}})- 2 \tau_4 A_{5\bar{5}} + A_{54} ( \bar{\lambda}_5 + \stackrel{\bar{5}}{\mu}_{5\bar{5}} ) + A_{44} (E_4 + i \chi \tau_4) - \frac{i \Lambda \chi}{4} (E_4 - 3 i \chi \tau_4) = 0.
\ee
Finally, $ijk = 55\bar{5}$ gives
\begin{align}
 A_{45} E_4 - B_{55\bar{5}}& - m_5^0(A_{5\bar{5}})- 2 A_{5\bar{5}} \lambda_5 \nonumber \\[1mm]
   & + i \chi \left[ 2 A_{54} E_4 + 2 G_5 - \Lambda E_5 + (A_{45} + 2 f_{45}) \tau_4 \right] + \chi^2 (2 A_{54} \tau_4 - \Lambda \lambda_5) = 0.
\end{align}

In summary, eqn.\ \eqref{11l} gives the following relations:
\begin{align}
& m_4^0(A_{45}) - m_5^0(A_{44}) = \tau_4 f_{45} + \lambda_5 A_{44} + A_{45} (\stackrel{5}{\mu}_{\bar{5}4} - 2 \tau_4) - \frac{\Lambda}{4} (E_5 - i \chi \lambda_5) - A_{54} (E_4 - i \chi \tau_4), \notag \\
 &m_4^0(f_{5\bar{5}}) - m_4^0(f_{\bar{5}5}) + 2 m_5^0 (A_{4\bar{5}})- 2 \mbar_5^0(A_{45}) = 2 (A_{5\bar{5}} - A_{\bar{5}5}) E_4  - 2 A_{4\bar{5}} ( 2 \lambda_5 + \stackrel{5}{\mu}_{\bar{5}5} ) + 2 A_{45} ( 2 \bar{\lambda}_5 + \stackrel{\bar{5}}{\mu}_{5\bar{5}} ) \nonumber \\[3mm]
   & \hspace{60mm} - 2 \tau_4 (f_{5\bar{5}} - f_{\bar{5}5}) + 2 i \chi \left[ 2 A_{44} E_4 + 2 G_4 - (A_{5\bar{5}} + A_{\bar{5}5} - \Lambda \chi^2) \tau_4 \right], \nonumber \\[3mm]
 & m_5^0(A_{5\bar{5}}) = - B_{55\bar{5}} + A_{45} E_4  - 2 A_{5\bar{5}} \lambda_5 + i \chi \left[ 2 A_{54} E_4 + 2 G_5 - \Lambda E_5 + (A_{45} + 2 f_{45}) \tau_4 \right]  \notag \\[3mm]
& \hspace{130mm} + \chi^2 (2 A_{54} \tau_4 - \Lambda \lambda_5),  \notag \\
&m_5^0(A_{54}) = B_{545} + A_{54} ( \stackrel{5}{\mu}_{\bar{5}5} - \lambda_5), \notag \\
& m_4^0(A_{5\bar{5}}) - \mbar_5^0(A_{54}) = - 2 \tau_4 A_{5\bar{5}} + A_{54} (\stackrel{\bar{5}}{\mu}_{5\bar{5}} + \bar{\lambda}_5 ) + A_{44} (E_4 + i \chi \tau_4) - \frac{i \Lambda \chi}{4} (E_4 - 3 i \chi \tau_4). \label{11l:summary}
\end{align} 

\subsection{Commutator C2$'$}

Previously we have considered commutators of GHP derivatives acting on an arbitrary GHP scalar $V_i$ of spin 1 with arbitrary boost weight $b$. The only such commutator that remains is $\left[\thop, \eth_i \right] V_j$, which is given by eqn.\ C2$'$ of Ref.\ \cite{ghp}. The $b$-dependent part of this expression is ({\textit{cf.}}\ eqn.\ (11c) of Ref.\ \cite{ricci})
\be
 \Delta L_{1i} - \delta_i L_{11} = L_{11} \left( L_{1i} - \tau_i \right) - \tau_j \rho'_{ji} + \rho_{ji} \kappa'_j - L_{1j} ( \rho'_{ji} + \stackrel{j}{M}_{i1} ) + \Psi'_i.
\label{11c}
\ee
For $i = 4$, we have
\be
 m_4^0(\lambda_1) - n^0(\tau_4) = 2 \tau_4 A_{44} + \left( \tau_4^2 - \frac{\Lambda}{4} \right) E_4,
\ee
while for $i = 5$ we obtain
\be
n^0(\lambda_5) - m_5^0(\lambda_1) = G_5  + \left(\frac{\Lambda}{4} - \tau_4^2 \right) E_5 + (\stackrel{5}{\mu}_{\bar{5}1} - \tau_4 E_4 + 2 i \chi \tau_4^2 ) \lambda_5 - 2  \tau_4 (A_{45} - i \chi A_{54}).
\ee

Now, we will investigate the boost-independent part of $\left[\thop, \eth_i \right] V_j$ ({\textit{cf.}}\ eqn.\ (11o) of Ref.\ \cite{ricci}),
\bea
 \Delta \stackrel{i}{M}_{jk} - \delta_k \stackrel{i}{M}_{j1} &=& \kappa'_j \rho_{ik} - \rho'_{jk} \tau_i + \tau_j \rho'_{ik} - \rho_{jk} \kappa'_i + \stackrel{i}{M}_{j1} (L_{1k} - \tau_k) \nonumber \\
 &\quad& + \stackrel{i}{M}_{l1} \stackrel{l}{M}_{jk} - \stackrel{i}{M}_{lk} \stackrel{l}{M}_{j1} - \stackrel{i}{M}_{jl} ( \rho'_{lk} + \stackrel{l}{M}_{k1} ) - \Psi'_{kij}.
\label{11o}
\eea
Here we have antisymmetry in $ij$, and hence the only independent components are $ijk = 454$, $455$, $45\bar{5}$, $5\bar{5}4$ and $5\bar{5}5$. The $ijk = 454$ component reduces to the following simple equation
\be
 m_4^0(A_{54}) = (\stackrel{5}{\mu}_{\bar{5}4} + \tau_4) A_{54}.
\ee
The $ijk = 455$ component is automatically satisfied using the expression for $B_{545}$ from eqn.\ \eqref{11l:summary}. The $45\bar{5}$ component gives
\bea
2 n^0(E_4) - 2 \mbar_5^0(A_{54}) + 2 i \chi  n^0(\tau_4) &=& - 2 G_4 - 2 \lambda_1 E_4 + 2 ( \stackrel{\bar{5}}{\mu}_{5\bar{5}} + \bar{\lambda}_5) A_{54} -\tau_4 \left( A_{5\bar{5}} + A_{\bar{5}5} + 2 E_4^2 \right) \nonumber \\[2mm]
   &\quad& - 2 i \chi  \left(\tau_4^2 + \frac{\Lambda}{4} \right) E_4 + \chi^2 \tau_4 \left( \Lambda + 2 \tau_4^2 \right).
\eea
Next, we consider $ijk = 5\bar{5}4$ to find
\be
 n^0 ( \stackrel{5}{\mu}_{\bar{5}4}) - m_4^0 ( \stackrel{5}{\mu}_{\bar{5}1} ) =  - (A_{44} +\tau_4 E_4) \stackrel{5}{\mu}_{\bar{5}4}.
\ee
Finally, the $5\bar{5}5$ component gives
\bea
 m_5^0 ( \stackrel{5}{\mu}_{\bar{5}1}) - n^0 ( \stackrel{5}{\mu}_{\bar{5}5} )  &=& G_5  + (f_{45} + A_{45} + i \chi A_{54} - i \chi \tau_4 \lambda_5 + \tau_4 E_5) \stackrel{5}{\mu}_{\bar{5}4} - (\stackrel{5}{\mu}_{\bar{5}5} + \lambda_5 ) \stackrel{5}{\mu}_{\bar{5}1} \nonumber \\[3mm]
    &\quad& + (E_4 - 2 i \chi \tau_4) A_{54}  +  \left[\lambda_1 - \tau_4 E_4 - i \chi \left( \frac{\Lambda}{4} + \tau_4^2 \right) \right] \stackrel{5}{\mu}_{\bar{5}5}.
\eea

In summary, from $\left[ \thop, \eth_i \right] V_j$ we obtain the following new relations:
\bea
 2 (n^0(E_4) - \mbar_5^0(A_{54}) +  i \chi  n^0(\tau_4)) &=& - 2 G_4 - 2 \lambda_1 E_4 + 2 ( \stackrel{\bar{5}}{\mu}_{5\bar{5}} + \bar{\lambda}_5) A_{54} -\tau_4 \left( A_{5\bar{5}} + A_{\bar{5}5} + 2 E_4^2 \right) \nonumber \\[2mm]
   &\quad& - 2 i \chi  \left(\tau_4^2 + \frac{\Lambda}{4} \right) E_4 + \chi^2 \tau_4 \left( \Lambda + 2 \tau_4^2 \right), \nonumber
\eea
\vspace*{-12mm}

\bea
m_4^0(\lambda_1) - n^0(\tau_4) &=& 2 \tau_4 A_{44} + \left( \tau_4^2 - \frac{\Lambda}{4} \right) E_4, \nonumber \\[1mm]
 n^0(\lambda_5) - m_5^0(\lambda_1) &=& G_5  + \left(\frac{\Lambda}{4} - \tau_4^2 \right) E_5 + (\stackrel{5}{\mu}_{\bar{5}1} - \tau_4 E_4 + 2 i \chi \tau_4^2 ) \lambda_5 - 2  \tau_4 (A_{45} - i \chi A_{54}), \nonumber \\[2mm]
 m_4^0(A_{54})  &=&(\stackrel{5}{\mu}_{\bar{5}4} + \tau_4) A_{54}, \nonumber \\[2mm]
 n^0 ( \stackrel{5}{\mu}_{\bar{5}4}) - m_4^0 ( \stackrel{5}{\mu}_{\bar{5}1} ) &=&  - (A_{44} +\tau_4 E_4) \stackrel{5}{\mu}_{\bar{5}4}, \nonumber \\[1mm]
 m_5^0 ( \stackrel{5}{\mu}_{\bar{5}1}) - n^0 ( \stackrel{5}{\mu}_{\bar{5}5} )  &=& G_5  + (f_{45} + A_{45} + i \chi A_{54} - i \chi \tau_4 \lambda_5 + \tau_4 E_5) \stackrel{5}{\mu}_{\bar{5}4} - (\stackrel{5}{\mu}_{\bar{5}5} + \lambda_5 ) \stackrel{5}{\mu}_{\bar{5}1} \nonumber \\[3mm]
    &\quad& + (E_4 - 2 i \chi \tau_4) A_{54}  +  \left[\lambda_1 - \tau_4 E_4 - i \chi \left( \frac{\Lambda}{4} + \tau_4^2 \right) \right] \stackrel{5}{\mu}_{\bar{5}5}.
\label{11o:summary}
\eea

\subsection{Commutators $\left[ n, m_i \right]$}

\label{sec:integrable}

Next, we consider the commutators
\be
 \left[ n, m_i \right] = - \kappa'_i\, \ell - (\tau_i - L_{1i})\, n + ( \stackrel{i}{M}_{j1} - \rho'_{ji} )\, m_j.
\label{n:mi:commutator}
\ee
We start with $i = 4$, for which eqn.\ \eqref{n:mi:commutator} reduces to
\be
 \left[ n, m_4 \right] = - \kappa'_4 \, \ell - \rho'_{44}\, m_4.
\label{n:m4:commutator}
\ee
Before doing an explicit calculation, recall eqns.\ \eqref{l:n:commutator} and \eqref{l:mi:commutator} with $i = 4$, which reduce to
\be
 \left[ \ell, n \right] = - L_{11}\, \ell - \tau_4\, m_4, \qquad \left[ \ell, m_4 \right] = - \tau_4\, \ell,
\ee
respectively. Together with eqn.\ \eqref{n:m4:commutator} above, these imply that the distribution spanned by $\{ \ell, n, m_4 \}$ is \emph{integrable}, i.e., tangent to three-dimensional submanifolds. 

Taking the $r$-component of \eqref{n:m4:commutator} gives the equation
\bea
 n^0(E_4) - m_4^0(E_1) &=& - G_4  + \tau_4 E_1 - (\lambda_1 + A_{44} + \tau_4 E_4) E_4 + \chi^2 \tau_4  \left( \frac{3\Lambda}{4} - \tau_4^2 \right).
\eea
On the other hand, from the $\mu$-components, we find the commutator
\be
 \left[ n^0, m_4^0 \right] = - (A_{44} + E_4 \tau_4)\, m_4^0.
\ee

Now, consider the commutator \eqref{n:mi:commutator} for $i = 5$,
\be
 \left[ n, m_5 \right] = - \kappa'_5\, \ell + L_{15}\, n - ( \stackrel{4}{M}_{51} + \rho'_{45} )\, m_4 + ( \stackrel{5}{M}_{\bar{5}1} - \rho'_{\bar{5}5} )\, m_5.
\label{n:m5:commutator}
\ee
Using previous results, we find that the $r$-component gives
\begin{align}
 n^0(E_5) - m_5^0(E_1) - i \chi m_5^0(\lambda_1) =& \ B_{55\bar{5}} + (A_{\bar{5}5} + i \chi \lambda_1 + E_1 - 5 \chi^2 \tau_4^2) \lambda_5 
 - (A_{45} - i \chi A_{54} + 2 E_5 \tau_4) E_4 \nonumber \\[2mm]
   & + \left[ \stackrel{5}{\mu}_{\bar{5}1}  - 2 \lambda_1 + i \chi\left( \frac{3 \Lambda}{4} - 2 \tau_4^2 \right) \right] E_5 - i \chi \tau_4 (f_{45} + 2 A_{45} - i \chi A_{54}).
\end{align}
For the $\mu$-components, we use the expression for $\left[ m_4^0, m_5^0 \right]$ coming from \eqref{mi:mj:commutator:summary:b} to find the commutator
\be
 \left[ n^0, m_5^0 \right] = \lambda_5\, n^0 - \left[ f_{45} + A_{45} + i \chi A_{54} + \tau_4 (E_5 - i \chi \lambda_5) \right] m_4^0 + \left[ i\chi\left(\frac{ \Lambda }{4} + \tau_4^2 \right) + \stackrel{5}{\mu}_{\bar{5}1} - \lambda_1 - \tau_4 E_4 \right] m_5^0.
\ee

In summary, then, the commutators \eqref{n:mi:commutator} give the relations
\begin{align}
&n^0(E_4) - m_4^0(E_1) = - G_4  + \tau_4 E_1 - (\lambda_1 + A_{44} + \tau_4 E_4) E_4 + \chi^2 \tau_4  \left( \frac{3\Lambda}{4} - \tau_4^2 \right), \notag \\[2mm]
 &n^0(E_5) - m_5^0(E_1) - i \chi m_5^0(\lambda_1) = \ B_{55\bar{5}} + (A_{\bar{5}5} + i \chi \lambda_1 + E_1 - 5 \chi^2 \tau_4^2) \lambda_5 
 - (A_{45} - i \chi A_{54} + 2 E_5 \tau_4) E_4 \nonumber \\[2mm]
   & \hspace{55mm} + \left[ \stackrel{5}{\mu}_{\bar{5}1}  - 2 \lambda_1 + i \chi\left( \frac{3 \Lambda}{4} - 2 \tau_4^2 \right) \right] E_5 - i \chi \tau_4 (f_{45} + 2 A_{45} - i \chi A_{54})
\label{n:mi:commutator:summary:a}
\end{align}
and the commutators
\begin{align}
 \left[ n^0, m_4^0 \right] &= - (A_{44} + E_4 \tau_4) m_4^0, \nonumber \\[2mm]
 \left[ n^0, m_5^0 \right] &= \lambda_5\, n^0 - \left[ f_{45} + A_{45} + i \chi A_{54} + \tau_4 (E_5 - i \chi \lambda_5) \right] m_4^0 + \left[ i\chi\left(\frac{ \Lambda }{4} + \tau_4^2 \right) + \stackrel{5}{\mu}_{\bar{5}1} - \lambda_1 - \tau_4 E_4 \right] m_5^0. 
\label{n:mi:commutator:summary:b}
\end{align}

\subsection{Equation NP1$'$}

The only Newman-Penrose equation yet to be investigated is NP1$'$ of Ref.\ \cite{ghp},
\be
 \Delta \rho'_{ij} - \delta_j \kappa'_i = - L_{11} \rho'_{ij} - (\tau_j  - 2 L_{1j} ) \kappa'_i + \stackrel{k}{M}_{ij} \kappa'_k - \stackrel{k}{M}_{i1} \rho'_{kj} - (\rho'_{kj} + \stackrel{k}{M}_{j1}) \rho'_{ik} - \Omega'_{ij}.
\label{11h}
\ee
With the results found above, we can use this to calculate $\Omega'_{ij}$, which are the only curvature components that we do not yet know. The results are quite complicated and will not be needed here, hence we will not display them. However, because $\Omega'_{ij}$ is symmetric and traceless, some components of this equation provide us with additional constraints.

It can be shown that all these constraints, with the exception of one are, in fact, implied by previously derived equations.  The new constraint comes from substituting $\Omega'_{44}$ calculated from the $ij = 44$ component of \eqref{11h} into $ij = 5\bar{5}$, using the tracefree condition $\Omega'_{5\bar{5}} = - \Omega'_{44}/2$. Thus, we obtain an important equation giving the derivative of $f_{5\bar{5}}$ along $n^0$:
\begin{align}
2 \bar{m}_5^0(B_{55\bar{5}}) + 2(\stackrel{\bar{5}}{\mu}_{5\bar{5}} + &3 \bar{\lambda}_5 ) B_{55\bar{5}}  + 2 n^0 (f_{\bar{5}5})  + (E_4 - i \chi \tau_4) m_4^0 (f_{\bar{5}5}) \nonumber \\[3mm] 
   & + 2 \left[ 3 \lambda_1 + 2 (2E_4 - i \chi \tau_4) \tau_4 \right] f_{\bar{5}5} + 2 (A_{4\bar{5}} + f_{4\bar{5}} - 2 i \chi A_{\bar{5}4}) f_{45} \,=\, 0.
\label{NP1prime}
\end{align}

\subsection{Additional algebraic constraint}

We can derive an additional algebraic constraint from the above equations by looking at the equations involving $f_{45}, f_{5\bar{5}}$ and the commutators $\left[ m_i^0, m_j^0 \right]$.

We begin by acting the commutator $\left[ m_4^0, m_5^0 \right]$ on $f_{4\bar{5}}$, using \eqref{A13:summary} to find
\be
 m_4^0 \left( m_5^0(f_{4\bar{5}}) \right) + 2 \tau_4\, m_5^0(f_{4\bar{5}}) + \left( m_5^0 ( \stackrel{5}{\mu}_{\bar{5}4} ) - A_{54} \right) f_{4\bar{5}} = 0.
\label{m40m50:f45b}
\ee
We now act $m_4^0$ on eqn.\ \eqref{A14:summary} and use \eqref{11p:summary} and eqn.\ \eqref{m40m50:f45b} to obtain
\be
 m_4^0 \left( m_4^0 (f_{5\bar{5}}) \right) + 6 \tau_4\, m_4^0 (f_{5\bar{5}}) + \left(\Lambda + 12 \tau_4^2 \right) f_{5\bar{5}} = 0.
\label{m40m40:f55b}
\ee
Next, consider the commutator $\left[ m_4^0, m_5^0 \right]$ acting on $f_{5\bar{5}}$, which gives
\be
 m_5^0 \left( m_4^0(f_{5\bar{5}}) \right) + 3 \lambda_5\, m_4^0(f_{5\bar{5}}) + 3 \left( m_4^0(\lambda_5) - \lambda_5 ( \stackrel{5}{\mu}_{\bar{5}4} - \tau_4 ) \right) f_{5\bar{5}} = 0.
\label{m50m40:f55b}
\ee
Now, we apply the commutator $\left[ m_4^0, m_5^0 \right]$ to $m_4^0(f_{5\bar{5}})$. After some algebra, this gives
\begin{align}
 &\Bigg[  m_4^0 \left( m_4^0(\lambda_5) \right) - 2 (\stackrel{5}{\mu}_{\bar{5}4} - 4 \tau_4 ) m_4^0(\lambda_5) 
 - \left( m_4^0 ( \stackrel{5}{\mu}_{\bar{5}4}) - (\stackrel{5}{\mu}_{\bar{5}4})^2 + 8 \tau_4 \stackrel{5}{\mu}_{\bar{5}4} -  8 \tau_4^2 - \frac{\Lambda}{4}\right) \lambda_5 + 8 \tau_4 A_{54} \Bigg] f_{5\bar{5}}
    \nonumber \\[4mm]
    & \hspace{115mm} - 2 A_{54} m_4^0(f_{5\bar{5}}) \, =\, 0. 
\label{derivation1}
\end{align}
By acting with $m_4^0$ on \eqref{m40:lambda5}, we obtain the relation
\bea
 m_4^0 \left( m_4^0(\lambda_5) \right) + ( \tau_4 - \stackrel{5}{\mu}_{\bar{5}4} ) m_4^0(\lambda_5)
 - \left[ m_4^0 ( \stackrel{5}{\mu}_{\bar{5}4} ) - \left( \tau_4^2 + \frac{\Lambda}{4} \right) \right] \lambda_5 &=& 0.
\eea
Together with \eqref{11p:summary}, this allows us to simplify \eqref{derivation1} to
\be
 \left( m_4^0(f_{5\bar{5}}) + 4 \tau_4 f_{5\bar{5}} \right) A_{54} = 0.
\label{derivation2}
\ee
Acting with $m_5^0$ and using eqn.\ \eqref{m50m40:f55b}, \eqref{11p:summary} and \eqref{11l:summary}, we find
\be
 \left(  m_4^0(f_{5\bar{5}}) + 4 \tau_4 f_{5\bar{5}} \right) B_{545} - 4 (A_{54})^2 f_{5\bar{5}} = 0.
\label{derivation3}
\ee
Now eqn.\ \eqref{derivation2} implies that either $A_{54} = 0$ or $m_4^0(f_{5\bar{5}}) = - 4 \tau_4 f_{5\bar{5}}$. In the former case, \eqref{11l:summary} implies that $B_{545} = 0$ and \eqref{derivation3} is automatically satisfied. In the latter case, we have $(A_{54})^2 f_{5\bar{5}} = 0$, which also holds when $A_{54} = 0$. Hence, both cases imply that
\be
 A_{54} f_{5\bar{5}} = 0.
\ee

\section{Coordinate Calculations}
\label{sec:coords}

\subsection{Introduction of coordinates}

In section 2 we determined the full $r$-dependence of the metric. However,  this process involved introducing many  functions of $x^\mu$ that are clearly not completely independent. In the previous section, after a thorough examination of the remaining Newman-Penrose equations, commutators of GHP derivatives and basis vectors, we were able to find several differential and algebraic constraints satisfied by these spacetime functions, which are summarised in appendix \ref{app:summary}. 

So far, we have been using coordinates $(r,x^\mu)$ where $r$ is an affine parameter along the null geodesics tangent to the multiple WAND, and $x^\mu$ are constant along these geodesics but otherwise arbitrary. In this section we will make a specific choice of $x^\mu$ adapted to the existence of the integrable submanifolds uncovered in section \ref{sec:integrable}. This will lead to a dramatic reduction in the number of independent functions and the number of equations constraining them, in comparison with the relations obtained in the previous section.

We start by observing that the results obtained so far allow us to write $\left( \d m_5 \right)_{ab} = 2 \stackrel{5}{M}_{[ba]}$ in the following form:
\be
 \d m_5  = \left[ \left( \rho'_{5\bar{5}} + \stackrel{5}{M}_{\bar{5}1} \right) \ell + \rho_{5\bar{5}} n + \left( \stackrel{4}{M}_{5\bar{5}} + \stackrel{5}{M}_{\bar{5}4} \right) m_4 + \stackrel{5}{M}_{\bar{5}5} \mbar_5 \right] \wedge m_5.
\ee
This implies that $m_5 \wedge \d m_5 = 0$, so there exist complex functions $\mathcal{W}$ and $z$ such that
\be
 m_5 = \bar{\mathcal{W}} \d \zbar.
\ee
Regard $x^\mu$ as coordinates on a four-dimensional manifold $\Sigma$. We can regard $m_4^0$, $m_5^0$ and $n^0$ as vector fields on $\Sigma$. From $\ell \cdot m_{\bar{5}} = 0$ we have $\partial_r z =0$ and, since $\d z \ne 0$, we must have $\partial_\mu z \ne 0$. It follows that we can use $z$ and $\bar{z}$ as coordinates on $\Sigma$. The condition $m_4 \cdot m_5 = 0$ implies that $m_4^0$ is tangent to surfaces of constant $z$ and similarly $n \cdot m_5 = 0$ implies that the same holds for $n^0$. Pick a three-dimensional surface $S \subset \Sigma$ which is transverse to $m_4^0$ and introduce a coordinate $u$ so that $(u,z,\bar{z})$ are coordinates on $S$. Now let $x$ be the parameter distance from $S$ along integral curves of $m_4^0$. This gives a coordinate chart $(u,x,z,\bar{z})$ on $\Sigma$ such that
\be
 m_4^0 = \frac{\partial}{\partial x}.
\ee
From the fact that $n^0$ is orthogonal to surfaces of constant $z$ we have
\be
 n^0 = N_u \frac{\partial}{\partial u} + N_x \frac{\partial}{\partial x}
\ee
with $N_u \ne 0$ and $N_x$ independent of $r$. Calculating the commutator $\left[ n^0,m_4^0 \right]$, eqn.\ \eqref{n0:m40:comm:summary}, in these coordinates gives two equations, the first being
\be
\partial_x N_u = 0.
\label{x-deriv:Nu}
\ee
We can then exploit the freedom $u \rightarrow u'(u,z,\zbar)$ to set $N_u = -1$. Hence we have
\be
 n^0 = - \frac{\partial}{\partial u} + N \frac{\partial}{\partial x}
\ee
for some real function $N$ (independent of $r$). The second equation coming from $\left[ n^0, m_4^0 \right]$ is then
\be
 A_{44} + E_4 \tau_4 = \partial_x N.
\ee
Since $m_5 \cdot m_5 = 0$, we have $(m_5^0)^{\zbar}=0$ so we can introduce the following notation for the components of $m_5^0$:
\be
 m_5^0 =  W \left( \frac{\partial}{\partial z} - L \frac{\partial}{\partial u} - Y \frac{\partial}{\partial x} \right),
\ee
where $W \ne 0$, $L$ and $Y$ are complex functions independent of $r$. Working out the basis as 1-forms gives
\bea
\label{basis1forms}
 \ell &=& - \d u - L \d z - \bar{L} \d \zbar, \\
 n &=& \d r - H \ell - (E_4 - r \tau_4) (\d x + Y \d z + \bar{Y} \d \zbar) \nonumber \\
   &\quad& \qquad + \frac{1}{W} \left[ \lambda_5 (r + i \chi) - E_5 \right] \d z + \frac{1}{\bar{W}} \left[ \bar{\lambda}_5 (r - i \chi) - \bar{E}_5 \right] \d \zbar, \\
 m_4 &=& - (N - r \tau_4) \ell + \d x + Y \d z + \bar{Y} \d \zbar, \\
 m_5 &=& \frac{r - i \chi}{\bar{W}} \d \zbar,
\eea
where
\be
 H = n^r - (E_4 - r \tau_4) (N - r \tau_4)
\ee
with $n^r$ given in (\ref{nr}). We see that the metric
\be
 g_{ab} = \ell_a n_b + n_a \ell_b + (m_4)_a (m_4)_b + (m_5)_a (\mbar_5)_b + (\mbar_5)_a (m_5)_b
\ee
depends on the functions just defined, $L$, $Y$, $N$, $W$, but also on the functions defined previously in a coordinate-independent way, in particular $\chi$, $\tau_4$,  $\lambda_5$, $E_4$, $E_5$, $E_1$, $\lambda_1$ and $f_{5\bar{5}}$ (the last three via $H$). Our aim now is to find coordinate expressions for these functions.

We proceed by examining the other commutators involving $m_4^0$, $m_5^0$ and $n^0$, which give coordinate expressions for some of the functions involved. From the commutator $\left[ m_4^0, m_5^0 \right]$, eqn.\ \eqref{m40:m50:comm:summary}, we find that
\be
 \partial_x L = \partial_x Y = 0,
\ee
hence $L = L(u,z,\zbar)$, $Y = Y(u,z,\zbar)$. Furthermore, we also find that
\be
 \partial_x W = W\left( \stackrel{5}{\mu}_{\bar{5}4} - \tau_4 \right).
\ee
By taking real and imaginary parts we obtain
\be
 \tau_4 = - \frac{1}{2} \left( \frac{\partial_x W}{W} + \frac{\partial_x \bar{W}}{\bar{W}} \right) = - \partial_x \ln |W|
\ee
and
\be
 \stackrel{5}{\mu}_{\bar{5}4} = \frac{1}{2} \left( \frac{\partial_x W}{W} - \frac{\partial_x \bar{W}}{\bar{W}} \right).
\ee
Next, calculating the commutator $\left[ n^0, m_5^0 \right]$, eqn.\ \eqref{n0:m50:comm:summary}, gives
\bea
 \lambda_5 &=& -W \partial_u L, \\
 \lambda_1 &=& \partial_u \ln |W| + (E_4 - N) \partial_x \ln |W|, \\
 \stackrel{5}{\mu}_{\bar{5}1} &=& - \frac{1}{2} \left( \frac{\partial_u W}{W} - \frac{\partial_u \bar{W}}{\bar{W}} \right) + \frac{N}{2} \left( \frac{\partial_x W}{W} - \frac{\partial_x \bar{W}}{\bar{W}} \right) - i \chi \left( \partial_x \ln |W| \right)^2 - \frac{i \Lambda \chi}{4},
\eea
together with
\be
 f_{45} + A_{45} + i \chi A_{54} + \tau_4 (E_5 - i \chi \lambda_5) = - W \left( \partial_u Y + N \partial_u L + Y \partial_x N - \partial N \right),
\label{n0:m50:comm:coords:d}
\ee
where we have used the differential operator
\be
 \partial \equiv \partial_z - L \partial_u.
\ee
Note that
\be
 \left[ \partial_x, \partial \right] = 0.
\ee
The only remaining commutator is $\left[ m_5^0, \mbar_5^0 \right]$, eqn.\ \eqref{m50:m50bar:comm:summary}, which gives
\bea
 |W|^2 \left( \bar{\partial} L - \partial \bar{L}  \right) &=& - 2 i \chi, \label{dL-dLbar} \\
 |W|^2 \left( \bar{\partial} Y - \partial \bar{Y}  \right) &=& 2 i \chi (N + E_4) \label{dY-dYbar}
\eea
and
\be
 \lambda_5 + \stackrel{5}{\mu}_{\bar{5}5} = - \frac{m_5^0(\bar{W})}{\bar{W}}.
\ee
This exhausts the information provided by the commutators, so we now turn to the differential and algebraic constraints obtained in the previous section.

We begin with eqn.\ \eqref{constraint03} for $m_4^0(\tau_4)$, which in our coordinates becomes
\be
 \partial_x^2 |W| + \frac{\Lambda |W|}{4} = 0.
\ee
If we exploit the freedom to shift $x$ by a function of the other coordinates, $x \rightarrow x' + f(u,z,\zbar)$, the general solution of this equation can be brought into the form\footnote{For example, if $\Lambda=0$ then we have $|W| = c_1 (u,z,\bar{z}) x + c_2(u,z,\bar{z})$. If $c_1 \ne 0$ then we can shift $x$ to eliminate $c_2$ so that $F=x$ and $P=c_1$. If $c_1=0$ then $F=1$ and $P=c_2$.}
\be
 |W| = F(x) P(u,z,\zbar),
\ee
with $F(x)$ and $P(u,z,\zbar)$ real and positive and $F$ satisfying the simple, one-dimensional ODE
\be
\label{Feq1a}
 F'' + \frac{\Lambda F}{4} = 0.
\ee
Notice that this implies
\be
 (F')^2 + \frac{\Lambda F^2}{4} \equiv \frac{\lambda}{3} = \mbox{constant}.
\label{lambda:def}
\ee
We then have
\be
 \tau_4 = - \frac{F'}{F}, \qquad \tau_4^2 + \frac{\Lambda}{4} = \frac{\lambda}{3 F^2}.
\ee
With these results we can use eqn.\ \eqref{constraint04} for $m_5^0(\tau_4)$ to determine $A_{54}$:
\be
 A_{54} = \frac{\lambda W Y}{3 F^2}.
\ee
We turn to eqn.\ \eqref{constraint01} for $m_4^0(\chi)$, which gives
\be
 \partial_x \left( \frac{\chi}{F^2} \right) = 0.
\ee
This implies that we can write
\be
 \chi = - F(x)^2 \, \Sigma(u,z,\zbar),
\ee
for some real function $\Sigma$ that does not depend on $x$.

Similarly, eqn.\ \eqref{constraint15} for $m_4^0(f_{45})$ implies that
\be
 \partial_x \left( \frac{f_{45}}{W} \right) = 0,
\ee
hence we write
\be
 f_{45} = W F_{45} (u,z,\zbar).
\ee
Using this result in eqn.\ \eqref{constraint16}, we find that
\be
 \partial_x \left( \frac{f_{5\bar{5}}}{F^4} \right) = \frac{j}{F^2},
\ee
where we have defined the complex function
\be
 j(u,z,\zbar) = 2 P^2 (\partial - \partial_u L) F_{4\bar{5}}.
\ee
We can solve for $f_{5\bar{5}}$ by writing
\be
 f_{5\bar{5}} = F(x)^4 \left[ G(x) j(u,z,\zbar) + k(u,z,\zbar) \right],
\ee
where $k(u,z,\zbar)$ is a complex function and $G(x)$ is such that
\be
 G'(x) = \frac{1}{F(x)^2}.
 \label{def:G}
\ee
Now consider eqn.\ \eqref{constraint14}. Substituting the results above, we have
\be
 - 4 Y F' (Gj + k) - \frac{Y j}{F} + F G \partial j + F \partial k = 3 F (Gj + k)\partial_u L.
\label{m50:f55b:coords}
\ee
Differentiating with respect to $x$, we find
\be
 \partial j = 3 j \partial_u L + \frac{2 F' Y j}{F} - \frac{4 \lambda Y}{3} (Gj + k).
\label{partial:j1}
\ee
Substituting this back into eqn.\ \eqref{m50:f55b:coords}, we obtain
\be
 \partial k = 3 k \partial_u L + \frac{4 F' Y k}{F} + \frac{Y j}{F^2} + \frac{2 G Y}{F} \left[ F' j + \frac{2 \lambda F}{3} (Gj + k) \right].
\label{partial:k1}
\ee
Differentiating \eqref{partial:j1} with respect to $x$ again, we find
\be
 \lambda Y j = 0.
\label{vanishing}
\ee
Furthermore, the algebraic condition \eqref{algebraic02} then implies
\be
 \lambda Y k = 0.
\label{vanishing2}
\ee
This means that we can rewrite \eqref{partial:j1} and \eqref{partial:k1} as follows:
\bea
 \partial j &=& 3 j \partial_u L + \frac{2 F' Y j}{F}, \label{partial:j} \\
 \partial k &=& 3 k \partial_u L + \frac{4 F' Y k}{F} + \frac{Y j}{F^2} + \frac{2 G F' Y j}{F}. \label{partial:k}
\eea

Thus, we have found the $x$-dependence of $f_{5\bar{5}}$ and partially constrained its dependence on the other coordinates via $j$ and $k$. We now systematically examine the remaining constraint equations to determine other functions of interest. In particular, we still need to find a coordinate expression for $E_5$ and $E_1$. We start with eqn.\ \eqref{constraint02}, which allows us to write
\be
 E_5 = i W F \left[ 2 \Sigma (F' Y + F \partial_u L) - F \partial \Sigma \right].
\ee
Next, eqn.\ \eqref{constraint05} determines the imaginary part of $A_{5\bar{5}}$:
\be
 A_{5\bar{5}} - A_{\bar{5}5} = 2 i F \left[ F' N \Sigma + F (\Sigma \partial_u \ln P - \partial_u \Sigma) \right].
\ee
With the real part given by \eqref{algebraic01},
\be
A_{5\bar{5}} + A_{\bar{5}5} = 2 E_1 + \frac{4}{3} \lambda F^2 \Sigma^2,
\ee
we find that
\be
 A_{5\bar{5}} = E_1 + \frac{2}{3} \lambda F^2 \Sigma^2 + i F \left[ F' N \Sigma + F (\Sigma \partial_u \ln P - \partial_u \Sigma) \right].
\ee
Eqn.\ \eqref{constraint08} gives
\be
 \partial_x (E_4 + N) = 0,
\ee
and, hence, one can write
\be
E_4 + N \equiv J(u,z,\zbar).
\ee
Using this in eqn.\ \eqref{constraint09}, we find that
\be
 A_{45} = W \left[ F_{45} - \frac{3 i \Lambda F^2 Y \Sigma}{4} - i F F' (\partial \Sigma - \Sigma \partial_u L) + \partial (N - J) - (N - J) \partial_u L - i (F')^2 Y \Sigma - Y \partial_x N \right].
\ee
Now, eqn.\ \eqref{n0:m50:comm:coords:d} simplifies to
\be
 F_{45} = \frac{2 i}{3} \lambda Y \Sigma + \frac{1}{2} \left( \partial J - J \partial_u L - \partial_u Y \right).
\ee
The next non-trivial equation is \eqref{constraint11}, which determines $E_1$
\be
 E_1 = \frac{1}{2} F^2 P^2 \left[ \bar{\partial} S + \partial \bar{S} - \frac{F'}{F} (\bar{\partial} Y + \partial \bar{Y}) \right] - \frac{\lambda P^2 |Y|^2}{3} - \frac{7 \lambda F^2 \Sigma^2}{6} + \frac{3 \Lambda F^4 \Sigma^2}{8} - \frac{1}{2} (N - J)^2,
\ee
where
\be
S(u,z,\zbar) \equiv \partial \ln P - \partial_u L.
\ee
Using this and previous results in eqn.\ \eqref{constraint13} determines the imaginary part of $f_{5\bar{5}}$:
\bea
 \frac{f_{5\bar{5}} - f_{\bar{5}5}}{F^4} &=& - 2 i P^2 \mathrm{Re} \left( \partial \bar{\partial} \Sigma - 2 \partial_u \bar{L} \partial \Sigma - \Sigma \partial_u \partial \bar{L} \right) - 2 i \Sigma P^2  (\bar{\partial} S + \partial \bar{S}) \nonumber \\
                                         &\quad& + \frac{8 i  F' P^2}{F} \mathrm{Re} \left[ \partial (\bar{Y} \Sigma) - \bar{Y} \Sigma \partial_u L \right] + 2 i \Lambda P^2 |Y|^2 \Sigma + \frac{8 i \lambda \Sigma^3}{3}.
\label{imaginary:f55b}
\eea

Let us now consider eqn.\ \eqref{constraint23}. Apart from $G_5$, we know explicitly all the quantities appearing in this expression, hence we can solve for $G_5$ to obtain
\be
 G_5 = \frac{i \lambda W}{3} \left( \frac{2 F' Y \Sigma}{F} + \partial \Sigma - \Sigma \partial_u L \right) + \frac{\lambda W Y}{3 F^2} (2 N - J) + \frac{W F'}{F} \partial_u Y - W \partial_u \left( \partial \ln P - \partial_u L \right).
\ee
Meanwhile, eqn.\ \eqref{constraint19} determines $B_{545}$, which is useful in computing $\Psi'_{545}$:
\be
 B_{545} = \frac{\lambda W^2}{3 F^2} \left( \partial Y + 2 Y S \right).
\label{B545:coords}
\ee
Similarly, eqn.\ \eqref{constraint21} determines $B_{55\bar{5}}$:
\bea
 B_{55\bar{5}} &=& (J - N + 3 i F F' \Sigma) W F_{45} - W F^2 P^2 \bar{\partial} \left( S^2 + \partial S - \frac{\Lambda Y^2}{4} \right) + \lambda W F^2 (\partial \Sigma^2 - 2 \Sigma^2 \partial_u L) \nonumber \\
               &\quad& + W F^2 P^2 (\bar{\partial} - \bar{Y} \partial_x) \left[ \frac{F'}{F} (\partial Y + 2 Y S) \right],
\eea
where we used previous results and the commutation relations
\be \label{commeqns}
 \left[ \partial, \bar{\partial} \right] = (\bar{\partial} L - \partial \bar{L})\ \partial_u, \qquad \left[ \partial, \partial_u \right] = \partial_u L\ \partial_u.
\ee
We can finally use this to rewrite the last constraint equation we need, eqn.\ \eqref{constraint30}, which is equivalent to a component of the Einstein equation that we have not yet incorporated into our results. After a lengthy calculation, \eqref{constraint30} can be put into the following form
\begin{align}
 &\partial_u \left( \frac{G j + k}{P^3} \right) - \frac{F'}{F} \frac{J}{P^3} (Gj + k) - \left( \frac{J}{F^2} - \frac{i F' \Sigma}{F} \right) \frac{j}{P^3} - \frac{2}{P} \left[ \frac{F_{45}}{F^2} - i \Lambda Y \Sigma - \frac{2 i F'}{F} (\partial \Sigma - \Sigma \partial_u L) \right] F_{4\bar{5}} \nonumber \\[3mm]
   & \hspace{30mm} + P \left( \partial + 2S - \frac{2 F' Y}{F} \right) \left[ \partial \left( \bar{S}^2 + \bar{\partial} \bar{S} - \frac{\Lambda \bar{Y}^2}{4} \right) - \frac{\lambda}{P^2} (\bar{\partial} \Sigma^2 - 2 \Sigma^2 \partial_u \bar{L}) \right] \nonumber \\[3mm]
   & \hspace{40mm} - P \left( \partial + 2S - Y \partial_x -  \frac{2 F' Y}{F} \right) (\partial - Y \partial_x) \left[ \frac{F'}{F} (\bar{\partial} \bar{Y} + 2 \bar{Y} \bar{S})  \right] = 0 .
\label{NP1prime:coords}
\end{align}
One can then verify that the remaining constraint equations written in appendix \ref{app:summary} are either automatically satisfied or determine some other functions that do not appear explicitly in the metric or in other calculations. For example, eqn.\ \eqref{constraint25} determines $G_4$, but we will not need this result.

\subsection{Summary of the results}

\label{sec:summary}

It is convenient to define a new ``radial" coordinate as
\be
R = \frac{r}{F(x)^2},
\ee
The general form of the metric is then
\bea
 ds^2 &=& \ell \otimes n + n \otimes \ell + m_4 \otimes m_4 + m_5 \otimes \mbar_5 + \mbar_5 \otimes m_5 \nonumber \\[3mm]
      &=& (\omega - J \ell)^2 + F^2 \left[ 2 \ell \left( \d R + U \d z + \bar{U} \d \zbar - \mathcal{H} \ell \right) + \frac{2 (R^2 + \Sigma^2)}{P^2} \d z \d \zbar \right],
\label{metric:coords}
\eea
where
\be
 \omega = \d x + Y \d z + \bar{Y} \d \zbar,
\ee
\be
 \ell = -\left(\d u + L \d z + \bar{L} \d \zbar \right),
\ee
\be
 U = i \partial \Sigma - \left(R + i \Sigma \right) \left(\partial_u L + \frac{2 F' Y}{F} \right)
\label{U:coords}
\ee
and
\bea
 \mathcal{H} &=& \frac{1}{F^2} \left[ H - \frac{1}{2} (N - r \tau_4)^2 + \frac{J^2}{2} \right] \nonumber \\
             &=& \frac{P^2}{2} (\partial \bar{S} + \bar{\partial} S) - \frac{F' P^2}{2 F} (\partial \bar{Y} + \bar{\partial} Y) - \frac{\lambda P^2 |Y|^2}{3 F^2} - \frac{\lambda R^2}{6} \nonumber \\
             &\quad& - R \left( \partial_u \ln P + \frac{F' J}{F} \right) + \frac{\mathrm{Re} (Gj + k) R + \mathrm{Im} (Gj + k) \Sigma}{R^2 + \Sigma^2}
\label{calH:coords}
\eea
with
\be
S = \partial \ln P - \partial_u L.
\label{S:def}
\ee
Note that the dependence of the metric on the radial coordinate $R$ and on the coordinate $x$ is completely determined. The $x$-dependence appears only via $F(x)$ and $G(x)$ where $F$ obeys \eqref{Feq1a}, with first integral \eqref{lambda:def}, and $G'(x)=1/F(x)^2$. Hence it is easy to determine the $x$-dependence explicitly for any given $\Lambda$ and $\lambda$. The metric is furthermore completely characterised by the complex functions $L(u,z,\zbar)$, $Y(u,z,\zbar)$ and $k(u,z,\zbar)$, and by the real functions $P(u,z,\zbar)$, $J(u,z,\zbar)$. The real functions $\Sigma(u,z,\zbar)$ and $J(u,z,\zbar)$ appearing in the metric are constrained by \eqref{dL-dLbar} and \eqref{dY-dYbar}:
\be
 P^2 (\bar{\partial} L - \partial \bar{L}) = 2 i \Sigma, \qquad P^2 (\bar{\partial} Y - \partial \bar{Y}) = - 2 i \Sigma J,
\label{dLY-dLYbar:coords}
\ee
while
\be
 j(u,z,\zbar) = 2 P^2 (\partial - \partial_u L) F_{4\bar{5}}, \qquad F_{45}(u,z,\zbar) = \frac{2 i}{3} \lambda Y \Sigma + \frac{1}{2} \left( \partial J - J \partial_u L - \partial_u Y \right).
\label{j:F45:coords}
\ee
The imaginary part of $k$ is determined by eqn.\ \eqref{imaginary:f55b}, which can be rewritten as
\bea
 \mathrm{Im} (Gj + k) &=& - P^2 \mathrm{Re} (\partial \bar{\partial} \Sigma - 2 \partial_u \bar{L} \partial \Sigma - \Sigma \partial_u \partial \bar{L}) - 2 \Sigma P^2 \mathrm{Re} (\partial \bar{S}) \nonumber \\[2mm]
                      &\quad& + \frac{4 F' P^2}{F} \mathrm{Re} \left[ \partial (\bar{Y} \Sigma) - \bar{Y} \Sigma \partial_u L \right] + \Lambda P^2 |Y|^2 \Sigma + \frac{4 \lambda \Sigma^3}{3}.
\label{imaginary:f55b-2}
\eea
And finally, these functions are related by the PDEs \eqref{partial:j}, \eqref{partial:k} and \eqref{NP1prime:coords}, repeated here for convenience:
\bea
 \partial j &=& 3 j \partial_u L + \frac{2 F' Y j}{F}, \label{partial:j-2} \\
 \partial k &=& 3 k \partial_u L + \frac{4 F' Y k}{F} + \frac{Y j}{F^2} + \frac{2 G F' Y j}{F}, \label{partial:k-2}
\eea
\begin{align}
 &\partial_u \left( \frac{G j + k}{P^3} \right) - \frac{F'}{F} \frac{J}{P^3} (Gj + k) - \left( \frac{J}{F^2} - \frac{i F' \Sigma}{F} \right) \frac{j}{P^3} - \frac{2}{P} \left[ \frac{F_{45}}{F^2} - i \Lambda Y \Sigma - \frac{2 i F'}{F} (\partial \Sigma - \Sigma \partial_u L) \right] F_{4\bar{5}} \nonumber \\[3mm]
   & \hspace{30mm} + P \left( \partial + 2S - \frac{2 F' Y}{F} \right) \left[ \partial \left( \bar{S}^2 + \bar{\partial} \bar{S} - \frac{\Lambda \bar{Y}^2}{4} \right) - \frac{\lambda}{P^2} (\bar{\partial} \Sigma^2 - 2 \Sigma^2 \partial_u \bar{L}) \right] \nonumber \\[3mm]
   & \hspace{40mm} - P \left( \partial + 2S - Y \partial_x -  \frac{2 F' Y}{F} \right) (\partial - Y \partial_x) \left[ \frac{F'}{F} (\bar{\partial} \bar{Y} + 2 \bar{Y} \bar{S})  \right] = 0,
\label{NP1prime:coords-2}
\end{align}
as well as the algebraic equations \eqref{vanishing} and \eqref{vanishing2}
\begin{equation}
 \lambda Y j = 0, \qquad \lambda Y k = 0.
\end{equation}
The form of these algebraic equations implies that the analysis can be divided into the following three cases: 1.\ $Y=0$, 2.\ $\lambda=0$ and 3.\ $\lambda \ne 0$, $j=k=0$. Before we examine these three cases separately, we will briefly discuss the residual coordinate freedom.

\subsection{Residual coordinate freedom} \label{sec:free}

The general solution has the following residual coordinate freedom\footnote{For $\lambda = \Lambda = 0,$ see case 2.1 below, we can choose $F=1$ so that $R=r$.}
\begin{equation}
 z \longrightarrow q(z), \qquad u \longrightarrow Q(u,z, \bar{z}), \qquad R \longrightarrow R/\partial_u Q
\end{equation}
in analogy with the residual coordinate freedom for four-dimensional (non-Kundt) algebraically special solutions \cite{exactsolutions, RRZ}.  Accordingly, the metric components transform as follows \cite{exactsolutions}:
\begin{gather}
 \Sigma \longrightarrow \Sigma/\partial_u Q, \qquad P \longrightarrow |\partial_z q| P/\partial_u Q, \qquad
 L \longrightarrow (\partial_u Q L - \partial_z Q)/\partial_z q
\end{gather}
as well as
\begin{gather}
 U \longrightarrow \frac{1}{\partial_z q\, \partial_u Q} \left( U + R \frac{\partial \partial_u Q}{\partial_u Q}\right), \qquad
 \mathcal{H} \longrightarrow \frac{1}{(\partial_u Q)^2} \left( \mathcal{H} + R \frac{\partial^2_u Q}{\partial_u Q}\right), \notag \\[3mm]
 Y \longrightarrow Y/\partial_z q, \qquad J \longrightarrow J/\partial_u Q, \qquad j \longrightarrow j/(\partial_u Q)^3,
\end{gather}
where we have used the fact that \cite{exactsolutions}
\begin{equation}
 \partial \longrightarrow (\partial_z q)^{-1} \partial, 
 \qquad S \longrightarrow \frac{1}{\partial_z q} \left( S + \frac{1}{2} \, \frac{\partial^2_z q}{\partial_z q}\right),
 \qquad k \longrightarrow k/(\partial_u Q)^3.
\end{equation}
The above freedom leads to a significant simplification when the multiple WAND is non-twisting and hence (because it is geodesic) hypersurface orthogonal. The vanishing twist condition $\rho_{[ij]}=0$ is equivalent to $\Sigma=0$. The first equation in \eqref{dLY-dLYbar:coords} reduces to
\be
 \partial \bar{L} - \bar{\partial} L = 0
\ee
and this is the integrability condition for the existence of a coordinate transformation of the above form that eliminates $L$. To see this, note that when $\ell$ is hypersurface orthogonal we must have $\ell = -X du'$ for some functions $X$ and $u'$. From the form of $\ell$ given in (\ref{basis1forms}) we deduce that $u'$ and $X$ must be independent of $r$ and $x$ with $X = (\partial u' /\partial u)^{-1}$ and $X \partial u'/\partial z = L$. A coordinate transformation as above with $Q = u'$ then transforms $L$ to zero. Hence in the hypersurface orthogonal case $\Sigma=0$ we can choose the coordinates so that $L=0$. 

\subsection{Case 1. $Y = 0$}

For $Y = 0$, the second equation in \eqref{dLY-dLYbar:coords} gives
\be
 J \Sigma = 0.
\ee
Therefore, this case naturally splits into two subcases.
\medskip

\noindent {\bf Case 1.1} $J = 0$. When $J = 0$, eqns.\ \eqref{j:F45:coords} imply
\be
 F_{45}=0, \qquad j = 0.
\ee
Because $j = 0$, eqn.\ \eqref{partial:j-2} is trivial; the remaining PDEs, eqns.\ \eqref{imaginary:f55b-2}, \eqref{partial:k-2} and \eqref{NP1prime:coords-2}, simplify to:
\be
 \mathrm{Im} (k) = - P^2 \mathrm{Re} (\partial \bar{\partial} \Sigma - 2 \partial_u \bar{L} \partial \Sigma - \Sigma \partial_u \partial \bar{L}) - 2 \Sigma P^2 \mathrm{Re} (\partial \bar{S}) + \frac{4 \lambda \Sigma^3}{3},
\ee
\be
 (\partial-3 \partial_u L) k =0
\ee
and
\be
 - \partial_u \left( \frac{k}{P^3} \right) = P \left( \partial + 2S \right) \partial (\bar{S}^2 + \bar{\partial} \bar{S}) - \lambda P \left( \partial + 2S \right) \left[ \frac{1}{P^2} (\bar{\partial} \Sigma^2 - 2 \Sigma^2 \partial_u \bar{L}) \right],
\ee
respectively.  But these are precisely the equations for algebraically special, twisting vacuum solutions in four dimensions with parameters $m + i M = -k$ and cosmological constant $\lambda$ \cite{exactsolutions, KaigorodovTimofeev1996}. In fact, the metric \eqref{metric:coords} reduces to
\be
 ds^2 = \d x^2 + F(x)^2 g_{(4)},
\label{metric:case1}
\ee
where $g_{(4)}$ is the four-dimensional metric
\be
 g_{(4)} = 2 \ell \left( \d R + U \d z + \bar{U} \d \zbar - \mathcal{H} \ell \right) + \frac{2 (R^2 + \Sigma^2)}{P^2} \d z \d \zbar
\ee
with
\be
 U = i \partial \Sigma - \left( R + i \Sigma \right) \partial_u L, \qquad \mathcal{H} = \frac{P^2}{2} (\partial \bar{S} + \bar{\partial} S) - \frac{\lambda R^2}{6} - R \partial_u \ln P - \frac{m R + M \Sigma}{R^2 + \Sigma^2}.
\ee
This is recognised as the general form of the metric for an algebraically special, twisting vacuum solution in four dimensions \cite{exactsolutions, KaigorodovTimofeev1996}. Therefore, our metric \eqref{metric:case1} is a warped product of such a four-dimensional metric $g_{(4)}$ with a fifth flat direction described by the coordinate $x$. These solutions are Class 1 of section \ref{sec:introsummary}. 

In this case, the only non-zero component of $\Phi_{ij}$ is
\be
 \Phi_{5\bar{5}} = \frac{k}{F^2 (R + i \Sigma)^3}.
\ee
Hence if $k \ne 0$ then the solution is genuinely type II. The solution will then be type III (or more special) if, and only if, $k = 0$. This is the same as the condition for $g_{(4)}$ to be type III (or more special) so such solutions are warped products of a flat direction with a 4d solution that is type III (or more special). When $k=0$, the only non-zero component of $\Psi'_{ijk}$ is
\be
 \Psi'_{55\bar{5}} = - \frac{W P^2}{F^2 (R - i \Sigma)^2} \left[ \bar{\partial} \left( S^2 + \partial S \right) - \frac{\lambda}{P^2} (\partial \Sigma^2 - 2 \Sigma^2 \partial_u L) \right].
\ee
Hence, the solution is type N (or conformally flat) if, and only if, the above expression also vanishes. 

If $g_{(4)}$ is type D then so is the 5d solution. Conversely, if a 5d solution is type D and (at least) one of the multiple WANDs has a rank 2 optical matrix then the solution must take the warped product form \eqref{metric:case1} with $g_{(4)}$ of type D \cite{wylleman}.

In the hypersurface orthogonal case we have $\Sigma=0$ and can choose $L=0$. These solutions are warped products of 4-dimensional \emph{Robinson-Trautman} (non-twisting) solutions with cosmological constant $\lambda$ with a fifth flat direction. Such solutions correspond to metric (84) in Ref.\ \cite{hyporthog}.
\medskip

\noindent {\bf Case 1.2} $\Sigma = 0$. When $\Sigma = 0$ we showed above that we can choose the coordinates so that $L = 0$. Using our general form \eqref{metric:coords} for the metric with $\Sigma = 0$, $L = Y = 0$, we find that
\be
 ds^2 = (\d x + J \d u)^2 + F^2 \left( -2 \mathcal{H} \d u^2 - 2 \d u \d R + \frac{2 R^2}{P^2} \d z \d \zbar \right),
\label{metric:ho}
\ee
where
\be
 2 \mathcal{H} = \Delta \ln P - \frac{\lambda R^2}{3} - 2 R \left( \partial_u \ln P + \frac{F'}{F} J \right) + \frac{2(Gj + k)}{R}
\label{calH:ho}
\ee
with $\Delta = 2 P^2 \partial_z \partial_{\zbar}$. Moreover, now we have $j = j(u)$ and $k = k(u)$ \emph{real} functions by virtue of eqns.\ \eqref{imaginary:f55b-2}--\eqref{partial:k-2}, and the relations between the various functions appearing in the metric simplify to
\be
 F_{45} = \frac{1}{2} \partial_z J, \qquad j = 2 P^2 \partial_z F_{4\bar{5}} = \frac{1}{2} \Delta J
\label{ho:eqn01}
\ee
and
\be
\partial_u \left( \frac{Gj + k}{P^3} \right) - \frac{F'}{F} \frac{J}{P^3} (Gj + k) - \frac{J}{F^2} \frac{j}{P^3} - \frac{2}{P} \frac{|F_{45}|^2}{F^2} + \frac{1}{4P^3} \Delta \Delta \ln P = 0.
\label{ho:eqn02}
\ee
Differentiating the above equation with respect to $x$, we obtain the additional relation
\be
\partial_u \left( \frac{j}{P^3} \right) + \frac{\lambda J}{3} \frac{(Gj + k)}{P^3} + J \frac{F'}{F} \frac{j}{P^3} + \frac{4 F'}{F} \frac{|F_{45}|^2}{P} = 0
\label{ho:eqn03}
\ee
and differentiating a second time with respect to $x$ we find that
\be
 \lambda |F_{45}|^2 = 0.
\ee
If $\lambda=0$ then the solution belongs to Case 2 so assume $\lambda \ne 0$ and hence $F_{45}=0$, for which $J = J(u)$ and $j = 0$. Eqn.\ \eqref{ho:eqn03} then gives
\be
 \lambda J k = 0.
\ee
Since $\lambda \ne 0$ we either have $J=0$, which is Case 1.1 (irrespective of whether or not $\Sigma$ vanishes) or $k=0$ which (since $j=0$) is Case 3. Therefore we have shown that all solutions in Case 1.2 are included in the other Cases and so we will not consider them further. 

\subsection{Case 2. $\lambda = 0$}

\label{sec:lambda0}

From eqn.\ \eqref{lambda:def}, we see that the case $\lambda = 0$ only arises when $\Lambda \leq 0$.  In this case, due to the special form of $F$, as found by solving eqn.\ \eqref{lambda:def}, there is a further residual coordinate freedom to shift $x$:
\begin{equation} \label{shiftx}
 x \longrightarrow x + f(u,z, \bar{z}),
\end{equation}
which we discuss further in the separate treatment of the cases $\Lambda = 0$ and $\Lambda < 0$ below.
\medskip

\noindent {\bf Case 2.1.} $\Lambda=0$. We then have $F={\rm constant}$ and we can absorb a constant into $P$ so that $F=1$, implying $G = x$.
The transformations induced on the metric components by the shift in $x$ \eqref{shiftx} are
\begin{equation} \label{2.1:shiftx}
 Y \longrightarrow Y - \partial f, \quad J \longrightarrow J - \partial_u f, 
 \qquad m \longrightarrow m + fj.
\end{equation}

By examining eqn.\ \eqref{imaginary:f55b-2}, we see that the left hand side contains a term proportional to $x$ but the right hand side is independent of $x$. Therefore, we must have
\be \label{j:real}
 \mathrm{Im}(j) = 0
\ee
and
\be
 \mathrm{Im} (k) = - P^2 \mathrm{Re} (\partial \bar{\partial} \Sigma - 2 \partial_u \bar{L} \partial \Sigma - \Sigma \partial_u \partial \bar{L}) - 2 \Sigma P^2 \mathrm{Re} (\partial \bar{S}).
\label{imaginary:f55b:case2p1}
\ee
In fact, $\mathrm{Im}(j)=0$ is not an extra condition and is automatically satisfied.  In particular, a simple calculation using the definition of $j$ \eqref{j:F45:coords} shows that it is real (see footnote 2).  Similarly, in eqn.\ \eqref{NP1prime:coords-2}, only the first term is $x$-dependent and hence we have
\be \label{uderivj}
 \partial_u \left( \frac{j}{P^3} \right) = 0.
\ee
The other terms are independent of $x$ and hence the resulting equation is
\be
 - \partial_u \left( \frac{k}{P^3} \right) =  - J \frac{j}{P^3} - \frac{2}{P} |F_{45}|^2 + P \left( \partial + 2S \right) \partial (\bar{S}^2 + \bar{\partial} \bar{S}).
\label{NP1prime:case2p1}
\ee
The remaining equations are \eqref{j:F45:coords}, \eqref{partial:j-2} and \eqref{partial:k-2}, which become, respectively,
\be
 j = 2 P^2 (\partial - \partial_u L) F_{4\bar{5}}, \qquad F_{45} = \frac{1}{2} \left( \partial J - J \partial_u L - \partial_u Y \right),
\label{j:F45:case2p1}
\ee
\be \label{derivj}
 \left( \partial - 3 \partial_u L \right) j = 0
\ee
and
\be
 \left( \partial - 3 \partial_u L \right) k = Y j.
\label{partial:k:case2p1}
\ee
Moreover, since $j$ is real, eqn.\ \eqref{j:real}, we can derive new equations by considering the integrability conditions on eqns.\ \eqref{uderivj} and \eqref{derivj} and its complex conjugate.  More specifically, we apply the operator equations in \eqref{commeqns} on $j$ and simplify the resulting expression using eqns.\ \eqref{uderivj} and \eqref{derivj}.  This gives the following two equations
\begin{gather} \label{intL0}
j \partial_u \left[ P (\partial \bar{L} - \bar{\partial}L) \right] = 0,  \qquad j \partial_u S = 0,
\end{gather}
i.e.
\begin{equation}
 j=0, \qquad \textup{or} \qquad \partial_u \left[ P (\partial \bar{L} - \bar{\partial}L) \right] = \partial_u S = 0.
\end{equation}
Thus, the analysis splits into two subcases.  Before discussing each of these cases separately, let us consider the non-zero components of the Weyl tensor in this case.

The non-zero components of the Weyl tensor with boost weight 0 are
\be
\label{bwzero}
 \Phi_{45} = \frac{W F_{45}}{(r - i \Sigma)^2}, \qquad \Phi_{5\bar{5}} = \frac{xj + k}{(r + i \Sigma)^3}.
\ee
Type III (or more special) solutions therefore must have $j = k = F_{45} = 0$. When these conditions hold, the only non-trivial component of the Weyl tensor with boost weight $-1$ is simply
\be
 \Psi'_{55\bar{5}} = - \frac{W P^2}{(r - i \Sigma)^2} \bar{\partial} \left( S^2 + \partial S \right).
\ee
\medskip

\noindent {\bf Case 2.1.1.} $j \neq 0$. Such solutions are strictly type II.  The metric is ($r = R$ when $F = 1$)
\be
 ds^2 = (\omega - J \ell)^2 + 2 \ell \left( \d r + U \d z + \bar{U} \d \zbar - \mathcal{H} \ell \right) + \frac{2 (r^2 + \Sigma^2)}{P^2} \d z \d \zbar
\label{metric:case2p1}
\ee
with 
\be
 U = i \partial \Sigma - \left( r + i \Sigma \right) \partial_u L
\ee
and
\be
  \mathcal{H} = \frac{P^2}{2} (\partial \bar{S} + \bar{\partial} S) - r \partial_u \ln P + \frac{[ x j + \mathrm{Re} (k) ] r + \mathrm{Im} (k) \Sigma}{r^2 + \Sigma^2}.
\ee
Additionally,
\begin{equation}
 \partial_u \left[ P (\partial \bar{L} - \bar{\partial}L) \right] = \partial_u S = 0.
\label{2.1.1:extra}
\end{equation}
These solutions give Class 2.1.1 of section \ref{sec:introsummary}. 

As a special case, consider the non-twisting case, $\Sigma = L = 0$.  In this case, $Y$ satisfies the integrability condition $\partial_z \bar{Y} = \partial_{\bar{z}} Y$, hence we can use the residual freedom \eqref{shiftx} to shift $x$ with an appropriately defined $f(u,z,\bar{z})$, see \eqref{2.1:shiftx}, to set $Y = 0$. If $j \ne 0$, then we must have $J \ne 0$, and eqns.\ \eqref{2.1.1:extra} give $P = P(z,\zbar)$. As explained in Ref.\ \cite{hyporthog}, the transformation $r = \alpha(u) r'$, $u = u(u')$ with $\d u/ \d u' = \alpha$ can be used to set $j = \mathrm{constant}$. One can also use the shift $x = x' + \beta(u)$, which gives that $k \longrightarrow k + \beta j$, to set $k = 0$. The resulting metric is
\be
 ds^2 = \left( \d x + J(u,z,\zbar) \d u \right)^2 - 2 \d u \d r - 2 \mathcal{H}(r,x,z,\zbar) \d u^2 + \frac{2 r^2}{P(z,\zbar)^2} \d z \d \zbar
\ee
with
\be
 2 \mathcal{H} = \Delta \ln P + \frac{2 x j}{r}.
\ee
The field equations \eqref{ho:eqn01} and \eqref{ho:eqn02} reduce to
\be
 \Delta J = 2 j = \mathrm{constant}, \qquad \Delta \Delta \ln P = 2 P^2 |\partial_z J|^2 + 4 j J.
\ee
This is the solution described by metric (94) in Ref.\ \cite{hyporthog}. 
\medskip

\noindent {\bf Case 2.1.2.} $j = 0$.  The metric is as in \eqref{metric:case2p1}, and now it is independent of $x$, with $\mathcal{H}$ given by
\be
 \mathcal{H} = \frac{P^2}{2} (\partial \bar{S} + \bar{\partial} S) - r \partial_u \ln P + \frac{\mathrm{Re} (k) r + \mathrm{Im} (k) \Sigma}{r^2 + \Sigma^2}.
\ee
Eqn.\ \eqref{imaginary:f55b:case2p1} is unchanged, but \eqref{NP1prime:case2p1} becomes
\be
 P^4 \left( \partial + 2S \right) \partial (\bar{S}^2 + \bar{\partial} \bar{S}) + P^3 \partial_u \left( \frac{k}{P^3} \right) = 2 P^2 |F_{45}|^2.
\ee
Furthermore, the inhomogeneous term on the right hand side of \eqref{partial:k:case2p1} vanishes, giving
\be
 \left( \partial - 3 \partial_u L \right) k = 0
\ee
and from \eqref{j:F45:case2p1} we obtain
\be
(\partial - \partial_u L) F_{4\bar{5}} = 0.
\ee
We identify this set of equations as the field equations governing an algebraically special Einstein-Maxwell system in four dimensions in the aligned, uncharged case ({\textit{cf.}}\ chapter 30 of Ref.\ \cite{exactsolutions}). Note that the metric \eqref{metric:case2p1} can be written in a Kaluza-Klein form
\be
 ds^2 = (\d x + \mathcal{A})^2 + g_{(4)},
\ee
where the Kaluza-Klein gauge field is
\be
 \mathcal{A} = Y \d z + \bar{Y} \d \zbar - J \ell.
\label{KKfield}
\ee
Here $g_{(4)}$ is a four-dimensional algebraically special metric that solves the Einstein-Maxwell system with vanishing cosmological constant in the aligned, uncharged case \cite{exactsolutions}. The corresponding Maxwell field is determined by the 1-form $\mathcal{A}$, with field strength
\be
 \d \mathcal{A} = 2 \ell \wedge (F_{45} \d z + F_{4\bar{5}} \d \zbar).
\label{KKfieldstrength}
\ee
Note that this is a Kaluza-Klein solution with a constant dilaton, which is possible because the Maxwell field is null. These solutions form Class 2.1.2 of section \ref{sec:introsummary}. 

In the non-twisting case $\Sigma = L = 0$ (and choosing $Y = 0$ as in Case 2.1.1), these solutions reduce to the metric (89) in Ref.\ \cite{hyporthog}, describing Kaluza-Klein uplifts of 4-dimensional algebraically special Robinson-Trautman metrics with a null Maxwell field to five dimensions.

From \eqref{bwzero} we see that the solution is type III (or more special) if, and only if, $k=0$ and $F_{45}=0$. But $F_{45}=0$ implies that $d \mathcal{A} = 0$ and hence $\mathcal{A}$ can be eliminated by a shift $x \rightarrow x + f(u,z,\zbar)$. The resulting solution is simply a product of a flat direction with a 4d Ricci flat metric of type III (or more special), i.e.\ it belongs to Class 1 of section \ref{sec:introsummary}.
\medskip

\noindent {\bf Case 2.2.} $\Lambda <0$. Let $\Lambda = -4/l^2$ with $l>0$. Then we have
\be
\left( \frac{F'}{F} \right)^2 = \frac{1}{l^2}
\ee
and we can arrange $F'/F = 1/l$ using the freedom $x \rightarrow - x$, $j \rightarrow - j$, $Y \rightarrow - Y$, $J \rightarrow - J$. More explicitly, we have, absorbing a constant into $P$,
\be
 F(x) = e^{x/l}
\ee
and
\be
 G(x) = - \frac{l}{2} e^{-2x/l} = - \frac{l}{2 F(x)^2}.
\ee
The transformations induced on the metric components by the shift in $x$ \eqref{shiftx} are as follows:
\begin{gather}
 \ell \longrightarrow \ell, \quad L \longrightarrow L, \qquad R \longrightarrow R e^{-2f/l} , \quad \Sigma \longrightarrow \Sigma e^{-2f/l},
 \quad P \longrightarrow P e^{-f/l}, \quad S \longrightarrow S - \frac{\partial f}{l},\notag \\[2mm]
 Y \longrightarrow Y - \partial f, \quad J \longrightarrow J - \partial_u f,  \quad
 m \longrightarrow e^{-4f/l} \left(m - \frac{lj}{2}\right), \quad M \longrightarrow M e^{-4f/l}, \quad j \longrightarrow j e^{-2f/l}\notag \\[2mm]
 U \longrightarrow U e^{-2f/l} + \frac{2}{l} R \partial f e^{-2f/l}, 
 \qquad \mathcal{H} \longrightarrow \mathcal{H} e^{-2f/l} + \frac{2}{l} R \partial_u f.
\end{gather}

Eqn.\ \eqref{partial:j-2} becomes
\be \label{2.2:dj}
 \partial j = 3 j \partial_u L + \frac{2 Y j}{l},
\ee
while eqn.\ \eqref{partial:k-2} reduces to
\be
 \partial k = 3 k \partial_u L + \frac{4 Y k}{l}
\ee
with $j$ as given in eqn.\ \eqref{j:F45:case2p1}.
Turning to eqn.\ \eqref{imaginary:f55b-2}, there are two $x$-dependent parts: terms proportional to $1/F^2$ and terms independent of $x$. These lead to two equations:
\be
 \mathrm{Im} (j) = 0
\ee
and
\be
 \mathrm{Im} (k) = - P^2 \mathrm{Re} \left[ \partial \bar{\partial} \Sigma - 2 \partial_u \bar{L} \partial \Sigma - \Sigma \partial_u \partial \bar{L} + 2 \Sigma \partial \bar{S} - \frac{4}{l} \partial (\bar{Y} \Sigma) + \frac{4}{l} \bar{Y} \Sigma \partial_u L + \frac{4 |Y|^2 \Sigma}{l^2} \right].
\ee
Note that, as before, the fact that $j$ is real does not imply any extra conditions---this can be shown from the definition of $j$ (see footnote 2).  A similar reasoning applied to eqn.\ \eqref{NP1prime:coords-2} gives
\be \label{2.2:udj}
 \frac{l}{2} \partial_u \left( \frac{j}{P^3} \right) + \frac{J j}{2 P^3} + \frac{2 |F_{45}|^2}{P} = 0
\ee
and
\begin{align}
\partial_u \left( \frac{k}{P^3} \right) - \frac{J k}{l P^3} + \frac{i \Sigma j}{l P^3} &+ \frac{4i}{l P} \left[ \partial \Sigma - \Sigma \left( \partial_u L + \frac{2 Y}{l} \right) \right] F_{4\bar{5}} \nonumber \\[2mm]
   & \quad + P \left[ \partial + 2 \left( S - \frac{Y}{l} \right) \right]  \partial \left[ \left( \bar{S} - \frac{\bar{Y}}{l} \right)^2 + \bar{\partial} \left( \bar{S} - \frac{\bar{Y}}{l} \right) \right] = 0.
\end{align}

As in the $\Lambda = 0$ case, the reality condition on $j$ allows us to derive two new equations from an integrability condition on eqns.\ \eqref{2.2:udj} and \eqref{2.2:dj} and its complex conjugate 
\begin{gather}
 3 j \partial_u \textup{ln}\left[ P (\partial \bar{L} - \bar{\partial}L) \right] - \frac{1}{l} \left(3 J j + P^2\, |\partial J - J \partial_u L - \partial_u Y|^2 \right) = 0, \\
 3 j \partial_u \left( S - \frac{Y}{l} \right) - \frac{j}{l} (\partial J - J \partial_u L - \partial_u Y|^2) 
 - \frac{P^2}{l} \left[ \partial - 2 \partial_u L + 2 \left( S - \frac{Y}{l} \right) \right] |\partial J - J \partial_u L - \partial_u Y|^2= 0.
\end{gather}
Note that the above equations reduce to those in \eqref{intL0} in the $l \longrightarrow \infty$ limit, as expected.

The metric is given by eqn.\ \eqref{metric:coords} with
\be
 U = i \partial \Sigma - (R + i \Sigma) \left( \partial_u L + \frac{2 Y}{l} \right),
\ee
\begin{align}
 \mathcal{H} = \frac{P^2}{2} \left[ \partial \left( \bar{S} - \frac{\bar{Y}}{l} \right) + \bar{\partial} \left( S - \frac{Y}{l} \right) \right] - R & \left( \partial_u \ln P + \frac{J}{l} \right) \notag \\[2mm]
 &+ \frac{1}{R^2 + \Sigma^2} \left\{ \left[ \mathrm{Re} (k) - \frac{l j}{2} e^{-2x/l} \right] R + \mathrm{Im} (k) \Sigma \right\}.
\end{align}
These solutions give Class 2.2 of section \ref{sec:introsummary}.

As a special case, when the multiple WAND $\ell$ is non-twisting, $\Sigma = L = 0$ (and as before using the residual freedom in shifting $x$ \eqref{shiftx} to set $Y = 0$), we recover the solution given in equation (100) of Ref.\ \cite{hyporthog}. 

Turning to the Weyl components, the non-zero components of $\Phi_{ij}$ are
\be
 \Phi_{45} = \frac{W F_{45}}{F^4(R - i \Sigma)^2}, \qquad \Phi_{5\bar{5}} = \frac{e^{-2x/l}}{(R + i \Sigma)^3} \left( k - \frac{lj}{2} e^{-2x/l} \right).
\ee
Similar to the previous case, these solutions are of type III (or more special) if, and only if, $k=0$ and $F_{45} = 0$ ($\implies j=0$). But the metric \eqref{metric:coords} contains again the term
\be
 \omega - J \ell = \d x + \mathcal{A},
\ee
with $\mathcal{A}$ as defined in \eqref{KKfield}. It turns out that eqn.~\eqref{KKfieldstrength} holds also in this case. Thus, with $F_{45} = 0$, we see that $\d \mathcal{A} = 0$ and $\mathcal{A}$ can be eliminated by a shift $x \rightarrow x + f(u,z,\zbar)$. The resulting solution has $Y = 0$ and $J = 0$, thus belonging to Case 1.1 above. It is then the warped product of a flat direction with a four-dimensional Ricci flat metric of type III (or more special), i.e.\ it belongs to the Class 1 solutions in our classification.

\subsection{Case 3. $\lambda \ne 0$, $j=k=0$.}

For $j = k = 0$, eqns.\ \eqref{partial:j-2} and \eqref{partial:k-2} are trivial, while \eqref{j:F45:coords} reduce to
\be
 (\partial - \partial_u L) F_{4\bar{5}} = 0, \qquad F_{45} = \frac{2 i}{3} \lambda Y \Sigma + \frac{1}{2} \left( \partial J - J \partial_u L - \partial_u Y \right).
\label{F45:case3}
\ee
Setting $j = k = 0$ in \eqref{imaginary:f55b-2} gives
\begin{align}
P^2 \mathrm{Re} (\partial \bar{\partial} \Sigma - 2 \partial_u \bar{L} \partial \Sigma - \Sigma \partial_u \partial \bar{L})& + 2 \Sigma P^2 \mathrm{Re} (\partial \bar{S}) \nonumber \\[2mm]
                      & - \frac{4 F' P^2}{F} \mathrm{Re} \left[ \partial (\bar{Y} \Sigma) - \bar{Y} \Sigma \partial_u L \right] - \Lambda P^2 |Y|^2 \Sigma - \frac{4 \lambda \Sigma^3}{3} = 0.
\label{reSigma}
\end{align}
Differentiating the above equation with respect to $x$ gives
\be
 \frac{4 \lambda P^2}{3 F^2} \mathrm{Re} \left[ \partial (\bar{Y} \Sigma) - \bar{Y} \Sigma \partial_u L \right] = 0
\ee
and, because we are assuming that $\lambda \ne 0$, we conclude that
\be
 \mathrm{Re} \left[ \partial (\bar{Y} \Sigma) - \bar{Y} \Sigma \partial_u L \right] = 0.
\label{imaginary:f55b:case3-1}
\ee
Hence, from eqn.~\eqref{reSigma}
\be
P^2 \mathrm{Re} (\partial \bar{\partial} \Sigma - 2 \partial_u \bar{L} \partial \Sigma - \Sigma \partial_u \partial \bar{L}) + 2 \Sigma P^2 \mathrm{Re} (\partial \bar{S}) - \Lambda P^2 |Y|^2 \Sigma - \frac{4 \lambda \Sigma^3}{3} = 0.
\label{imaginary:f55b:case3-2}
\ee
Eqn.\ \eqref{NP1prime:coords-2} can be rewritten to take the following form:
\begin{align}
 &- \frac{2}{P} \left[ \frac{F_{45}}{F^2} - i \Lambda Y \Sigma - \frac{2 i F'}{F} (\partial \Sigma - \Sigma \partial_u L) \right] F_{4\bar{5}} + P \left( \partial + 2S - \frac{2 F' Y}{F} \right) \partial \left( \bar{S}^2 + \bar{\partial} \bar{S} - \frac{\Lambda \bar{Y}^2}{4} \right) \nonumber \\[3mm]
   & \hspace{30mm} - \lambda P \left( \partial + 2S - \frac{2 F' Y}{F} \right) \left[ \frac{1}{P^2} (\bar{\partial} \Sigma^2 - 2 \Sigma^2 \partial_u \bar{L}) \right]  - \frac{\lambda P}{3 F^2} |\partial Y + 2 Y S|^2\nonumber \\[3mm]
	 & \hspace{65mm} - P \left[ \frac{F'}{F} (\partial + 2S) + \frac{\Lambda Y}{2} \right] \partial \left( \bar{\partial} \bar{Y} + 2 \bar{Y} \bar{S} \right) = 0.
\label{NP1prime:case3}
\end{align}
Taking the derivative of the above equation with respect to $x$ gives
\begin{align}
&\frac{4}{F^2 P} \left[ \frac{F'}{F} F_{45} - \frac{i \lambda}{3} (\partial \Sigma - \Sigma \partial_u L) \right] F_{4\bar{5}} + \frac{2 \lambda P Y}{3 F^2} \partial \left( \bar{S}^2 + \bar{\partial} \bar{S} - \frac{\Lambda \bar{Y}^2}{4} \right) - \frac{2 \lambda^2 Y}{3 F^2 P} \left( \bar{\partial} \Sigma^2 - 2 \Sigma^2 \partial_u \bar{L} \right) \nonumber \\[3mm]
 & \hspace{50mm} + \frac{\lambda P}{3 F^2} \left[ (\partial + 2 S) \partial (\bar{\partial} \bar{Y} + 2 \bar{Y} \bar{S}) + \frac{2 F'}{F} |\partial Y + 2 Y S|^2 \right] = 0.
\end{align}
Multiplying the above equation by $F^2$ and differentiating with respect to $x$ again gives
\be
 - \frac{2 \lambda}{3 F^2 P^2} \left( 2 |F_{45}|^2 + \frac{\lambda P^2}{3} |\partial Y + 2 Y S|^2 \right) = 0.
\ee
Again, since we are assuming that $\lambda \ne 0$, we conclude that the term in brackets is zero,
\be
 |F_{45}|^2 + \frac{\lambda P^2}{6} |\partial Y + 2 Y S|^2 = 0.
\label{vanishing:case3}
\ee
Substituting this result back into \eqref{NP1prime:case3} gives an equation with 2 $x$-dependent parts: terms proportional to $F'/F$ and terms independent of $x$. Because $\partial_x (F'/F) = -\lambda/(3 F^2) \ne 0$, we then have two equations:
\be
\left( \partial + 2 S \right) \left[ \partial \left( \bar{S}^2 + \bar{\partial} \bar{S} - \frac{\Lambda \bar{Y}^2}{4} \right) - \frac{\lambda}{P^2} \left( \bar{\partial} \Sigma^2 - 2 \Sigma^2 \partial_u \bar{L} \right) \right] = \frac{\Lambda Y}{2} \left[ \partial \left( \bar{\partial} \bar{Y} + 2 \bar{Y} \bar{S} \right) - \frac{4 i \Sigma F_{4\bar{5}}}{P^2} \right]
\label{NP1prime:case3-1}
\ee
and
\begin{align}
 \left( \partial + 2 S \right) \partial \left( \bar{\partial} \bar{Y} + 2 \bar{Y} \bar{S} \right) =  &- 2 Y \left[ \partial \left( \bar{S}^2 + \bar{\partial} \bar{S} - \frac{\Lambda \bar{Y}^2}{4} \right) - \frac{\lambda}{P^2} \left( \bar{\partial} \Sigma^2 - 2 \Sigma^2 \partial_u \bar{L} \right) \right] \notag \\[2mm]
 & \hspace{70mm} + \frac{4 i F_{4\bar{5}}}{P^2} \left( \partial \Sigma - \Sigma \partial_u L \right).
\label{NP1prime:case3-2}
\end{align}
The two equations above can be written more compactly as
\begin{gather}
 (\partial + 2 S)\, \bar{{\Xi}}_1 - \frac{\Lambda}{2}\, Y\, \bar{{\Xi}}_2 = 0, \label{NP1prime:case3a} \\[2mm]
 (\partial + 2 S)\, \bar{{\Xi}}_2 + 2\, Y\, \bar{{\Xi}}_1 = 0 \label{NP1prime:case3b}
\end{gather}
with
\begin{gather}
 \Xi_1 = \bar{\partial} \left( S^2 + \partial S - \frac{\Lambda Y^2}{4} \right) - \frac{\lambda}{P^2} \left( \partial \Sigma^2 - 2 \Sigma^2 \partial_u L \right), \label{def:Xi1} \\[2mm]
 \Xi_2 = \bar{\partial} \left( \partial Y + 2 Y S \right) + \frac{4 i \Sigma F_{45}}{P^2}.
 \label{def:Xi2}
\end{gather}
The metric is given by eqn.\ \eqref{metric:coords} with $U$ given by the same expression as in \eqref{U:coords}, but with $\mathcal{H}$ now given by
\be
 \mathcal{H} = \frac{P^2}{2} (\partial \bar{S} + \bar{\partial} S) - \frac{F' P^2}{2 F} (\partial \bar{Y} + \bar{\partial} Y) - \frac{\lambda P^2 |Y|^2}{3 F^2} - \frac{\lambda R^2}{6} - R \left( \partial_u \ln P + \frac{F' J}{F} \right).
\ee
These solutions form Class 3 of section \ref{sec:introsummary}. 

The only non-zero component of the Weyl tensor with boost weight 0 in this case is
\be
 \Phi_{45} = \frac{W F_{45}}{F^4 (R - i \Sigma)^3}.
\label{Phi45:case3}
\ee
Hence the solution is type III (or more special) if, and only if, $F_{45}=0$. For such solutions, equation \eqref{vanishing:case3} reduces to
\be
\label{type3}
\partial Y + 2 Y S = 0.
\ee
In fact, with these restrictions, we also have that $\Psi'_{445} = \Psi'_{54\bar{5}} = \Psi'_{545} = 0$, the latter because $B_{545} = 0$ from eqn.\ \eqref{B545:coords}. Therefore, among the boost weight $-1$ components of the Weyl tensor, the only nontrivial component is
\be
 \Psi'_{55\bar{5}} = - \frac{W P^2\, \Xi_1}{F^2 (R - i \Sigma)^2}
\ee
with $\Xi_1$ as defined in eqn.\ \eqref{def:Xi1}.  But, in this case $\Xi_2=0$, see eqn.~\eqref{def:Xi2}. When $Y\neq 0$, eqns.\ \eqref{NP1prime:case3a} and \eqref{NP1prime:case3b} reduce to
\be
\Xi_1 = 0,
 \label{3:N}
\ee
which implies that $\Psi'_{55\bar{5}} = 0$. Therefore, such solutions are actually \emph{type N} (or conformally flat), i.e., we have shown that there are no genuinely type III solutions with $Y\neq0$ in this case. These type N (or conformally flat) solutions are characterised by eqns.\ \eqref{imaginary:f55b:case3-1} and \eqref{imaginary:f55b:case3-2}, which remain unchanged, and \eqref{F45:case3}, which becomes
\be
 \partial J - J \partial_u L - \partial_u Y + \frac{4 i}{3} \lambda Y \Sigma = 0,
\ee
as well as \eqref{type3} and \eqref{3:N}.  It can be shown by analysing the components of $\Omega_{ij}'$ that such solutions are not necessary conformally flat.  In particular, $\Omega'_{55} \neq 0.$

Note that when $\lambda>0$, eqn.\ \eqref{vanishing:case3} implies that $F_{45}=0$. Hence, any $\lambda>0$, $Y\neq0$ solution in this case is type N (or conformally flat).

Finally, consider as a special case the non-twisting case $\Sigma=L=0.$  Such solutions correspond to the solutions given by eqns.\ (102) and (103) of Ref.\ \cite{hyporthog}.  In particular, eqns.\ (102) and (103) of Ref.\ \cite{hyporthog} are equivalent to eqns.\ \eqref{F45:case3} and \eqref{NP1prime:case3} above, respectively.  Note that the $x$-dependence (or $y$-dependence in the language of that paper) of eqn.\ (103) of Ref.\ \cite{hyporthog} is not analysed fully there, which is why there are no equations there analogous to eqns.\ \eqref{vanishing:case3}, \eqref{NP1prime:case3a} and \eqref{NP1prime:case3b} above.

\appendix

\renewcommand{\theequation}{A.\arabic{equation}}
\setcounter{equation}{0}
\section{Summary}
\label{app:summary}

The equations constraining the $r$-independent functions derived in sections \ref{sec:constraint} are summarised in this appendix for convenience.

\subsection*{Differential relations}

\be
 n^0(\chi) = \frac{i(A_{5\bar{5}} - A_{\bar{5}5})}{2} - \chi (\lambda_1 + \tau_4 E_4)
\label{constraint05}
\ee
\be
 m_4^0(\chi) = - 2 \tau_4 \chi
\label{constraint01}
\ee
\be
 m_5^0(\chi) = - i E_5 - 2 \chi \lambda_5
\label{constraint02}
\ee
\be
 m_4^0(\tau_4) = \tau_4^2 + \frac{\Lambda}{4}
\label{constraint03}
\ee
\be
 m_5^0(\tau_4) = - A_{54}
\label{constraint04}
\ee
\be
m_4^0(\lambda_5) = (\stackrel{5}{\mu}_{\bar{5}4} - \tau_4) \lambda_5
\label{constraint06}
\ee
\be
 m_5^0(\bar{\lambda}_5) - \mbar_5^0(\lambda_5) = 2 i \chi \lambda_1 + 2 i \chi \tau_4 E_4 + \lambda_5 \stackrel{\bar{5}}{\mu}_{5\bar{5}} - \bar{\lambda}_5 \stackrel{5}{\mu}_{\bar{5}5} - A_{5\bar{5}} + A_{\bar{5}5}
\label{constraint07}
\ee
\be
 m_4^0(E_4) = - A_{44} - \tau_4 E_4
\label{constraint08}
\ee
\be
 m_5^0(E_4) = - A_{45} + f_{45} - \lambda_5 E_4 - \tau_4 E_5 + i \chi (3 A_{54} + \lambda_5 \tau_4)
\label{constraint09}
\ee
\be
m_4^0 ( \stackrel{5}{\mu}_{\bar{5}5} ) - m_5^0 ( \stackrel{5}{\mu}_{\bar{5}4} ) =  - A_{54} 
+ ( \stackrel{5}{\mu}_{\bar{5}4} - \tau_4) \stackrel{5}{\mu}_{\bar{5}5}
\label{constraint10}
\ee
\begin{align}
 m_5^0 ( \stackrel{\bar{5}}{\mu}_{5\bar{5}} ) + \mbar_5^0 ( \stackrel{5}{\mu}_{\bar{5}5} ) = - A_{5\bar{5}} - A_{\bar{5}5} & - E_4^2 - \lambda_5 \stackrel{\bar{5}}{\mu}_{5\bar{5}} - \bar{\lambda}_5 \stackrel{5}{\mu}_{\bar{5}5} \nonumber \\[2mm]
   & - 2 \stackrel{5}{\mu}_{\bar{5}5} \stackrel{\bar{5}}{\mu}_{5\bar{5}} - 2 i \chi ( \stackrel{5}{\mu}_{\bar{5}1} + E_4 \stackrel{5}{\mu}_{\bar{5}4} ) + \left( \frac{\Lambda}{2} - \tau_4^2 \right) \chi^2
\label{constraint11}
\end{align}
\be
 m_4^0(E_5) = 2 i \chi A_{54}  - (3 \tau_4 - \stackrel{5}{\mu}_{\bar{5}4}) E_5 
\label{constraint12}
\ee
\begin{align}
\mbar_5^0(E_5) - m_5^0(\bar{E}_5) - i \chi \left[m_5^0(\bar{\lambda}_5) + \mbar_5^0(\lambda_5)\right] =&  \lambda_5 \bar{E}_5 - \bar{\lambda}_5 E_5 + f_{5\bar{5}} - f_{\bar{5}5} - E_5 \stackrel{\bar{5}}{\mu}_{5\bar{5}} + \bar{E}_5 \stackrel{5}{\mu}_{\bar{5}5} \nonumber \\
   &\  + i \chi (\lambda_5 \stackrel{\bar{5}}{\mu}_{5\bar{5}} + \bar{\lambda}_5 \stackrel{5}{\mu}_{\bar{5}5} - 4 E_1 -2 E_4^2 -6 \chi^2 \tau_4^2)
\label{constraint13}
\end{align}
\be
 m_5^0(f_{5\bar{5}}) = - 3 \lambda_5 f_{5\bar{5}}
\label{constraint14}
\ee
\be
 m_4^0(f_{45}) = (\stackrel{5}{\mu}_{\bar{5}4} - \tau_4 ) f_{45}
\label{constraint15}
\ee
\be
 2 m_5^0(f_{4\bar{5}}) - m_4^0(f_{5\bar{5}})=  - 2 ( 2 \lambda_5 + \stackrel{5}{\mu}_{\bar{5}5} ) f_{4\bar{5}} + 4 \tau_4 f_{5\bar{5}} 
\label{constraint16}
\ee
\be
  m_4^0(A_{45}) - m_5^0(A_{44}) = \tau_4 f_{45} + \lambda_5 A_{44} + (\stackrel{5}{\mu}_{\bar{5}4} - 2 \tau_4) A_{45} - \frac{\Lambda}{4} (E_5 - i \chi \lambda_5) - (E_4 - i \chi \tau_4) A_{54}
\label{constraint17}
\ee
\begin{align}
 m_4^0(f_{5\bar{5}}) - m_4^0(f_{\bar{5}5}) + 2 m_5^0 (A_{4\bar{5}}) - 2& \mbar_5^0(A_{45}) = 2 (A_{5\bar{5}} - A_{\bar{5}5}) E_4  - 2 A_{4\bar{5}} ( 2 \lambda_5 + \stackrel{5}{\mu}_{\bar{5}5} ) + 2 A_{45} ( 2 \bar{\lambda}_5 + \stackrel{\bar{5}}{\mu}_{5\bar{5}} ) 
 \nonumber \\
   &   - 2 \tau_4 (f_{5\bar{5}} - f_{\bar{5}5})  + 2 i \chi \left[ 2 A_{44} E_4 + 2 G_4 - (A_{5\bar{5}} + A_{\bar{5}5} - \Lambda \chi^2) \tau_4 \right]
\label{constraint18}
\end{align}
\be
  m_5^0(A_{54}) = B_{545} + A_{54} (\stackrel{5}{\mu}_{\bar{5}5} - \lambda_5 )
\label{constraint19}
\ee
\be
 m_4^0(A_{5\bar{5}}) - \mbar_5^0(A_{54}) =  - 2 \tau_4 A_{5\bar{5}} + ( \bar{\lambda}_5 + \stackrel{\bar{5}}{\mu}_{5\bar{5}} ) A_{54} + (E_4 + i \chi \tau_4) A_{44} - \frac{i \Lambda \chi}{4} (E_4 - 3 i \chi \tau_4)
\label{constraint20}
\ee
\begin{align}
 m_5^0(A_{5\bar{5}}) = - B_{55\bar{5}} + A_{45} E_4  - 2 A_{5\bar{5}} \lambda_5 + i \chi \left[ 2 A_{54} E_4 + 2 G_5 - \Lambda E_5 + (A_{45} + 2 f_{45}) \tau_4 \right]& \nonumber \\
   & \hspace{-10mm} + \chi^2 (2 A_{54} \tau_4 - \Lambda \lambda_5)
\label{constraint21}
\end{align}
\be
  m_4^0(\lambda_1) - n^0(\tau_4) =   2 \tau_4 A_{44} + \left( \tau_4^2 - \frac{\Lambda}{4} \right) E_4
\label{constraint22}
\ee
\be
 n^0(\lambda_5) - m_5^0(\lambda_1)  = G_5  + \left(\frac{\Lambda}{4} - \tau_4^2 \right) E_5 + (\stackrel{5}{\mu}_{\bar{5}1} - \tau_4 E_4 + 2 i \chi \tau_4^2 ) \lambda_5 - 2  \tau_4 (A_{45} - i \chi A_{54})
\label{constraint23}
\ee
\be
 m_4^0(A_{54}) = (\stackrel{5}{\mu}_{\bar{5}4} + \tau_4) A_{54}
\label{constraint24}
\ee
\begin{align}
2 (n^0(E_4) - \mbar_5^0(A_{54}) +  i \chi  n^0(\tau_4)) = - 2 G_4 - 2 \lambda_1 E_4 +& 2 ( \stackrel{\bar{5}}{\mu}_{5\bar{5}} + \bar{\lambda}_5) A_{54} -\tau_4 \left( A_{5\bar{5}} + A_{\bar{5}5} + 2 E_4^2 \right) \nonumber \\[2mm]
   & - 2 i \chi  \left(\tau_4^2 + \frac{\Lambda}{4} \right) E_4 + \chi^2 \tau_4 \left( \Lambda + 2 \tau_4^2 \right)
\label{constraint25}
\end{align}
\be
 n^0( \stackrel{5}{\mu}_{\bar{5}4} ) - m_4^0 ( \stackrel{5}{\mu}_{\bar{5}1} ) =   - (A_{44} + \tau_4 E_4) \stackrel{5}{\mu}_{\bar{5}4}
\label{constraint26}
\ee
\begin{align}
 m_5^0 ( \stackrel{5}{\mu}_{\bar{5}1}) - n^0 ( \stackrel{5}{\mu}_{\bar{5}5} ) = G_5  + &(f_{45} + A_{45} + i \chi A_{54} - i \chi \tau_4 \lambda_5 + \tau_4 E_5) \stackrel{5}{\mu}_{\bar{5}4} - (\stackrel{5}{\mu}_{\bar{5}5} + \lambda_5 ) \stackrel{5}{\mu}_{\bar{5}1} \nonumber \\[3mm]
    & + (E_4 - 2 i \chi \tau_4) A_{54}  +  \left[\lambda_1 - \tau_4 E_4 - i \chi \left( \frac{\Lambda}{4} + \tau_4^2 \right) \right] \stackrel{5}{\mu}_{\bar{5}5}.
\label{constraint27}
\end{align}
\be
 n^0(E_4) - m_4^0(E_1) = - G_4  + \tau_4 E_1 - (\lambda_1 + A_{44} + \tau_4 E_4) E_4 + \chi^2 \tau_4  \left( \frac{3\Lambda}{4} - \tau_4^2 \right)
\label{constraint28}
\ee
\begin{align}
&n^0(E_5) - m_5^0(E_1) - i \chi m_5^0(\lambda_1) = \ B_{55\bar{5}} + (A_{\bar{5}5} + i \chi \lambda_1 + E_1 - 5 \chi^2 \tau_4^2) \lambda_5 
 - (A_{45} - i \chi A_{54} + 2 E_5 \tau_4) E_4 \nonumber \\[2mm]
   & \hspace{55mm} + \left[ \stackrel{5}{\mu}_{\bar{5}1}  - 2 \lambda_1 + i \chi\left( \frac{3 \Lambda}{4} - 2 \tau_4^2 \right) \right] E_5 - i \chi \tau_4 (f_{45} + 2 A_{45} - i \chi A_{54})
\label{constraint29}
\end{align}
\begin{align}
2 \bar{m}_5^0(B_{55\bar{5}}) + 2(\stackrel{\bar{5}}{\mu}_{5\bar{5}} + &3 \bar{\lambda}_5 ) B_{55\bar{5}}  + 2 n^0 (f_{\bar{5}5})  + (E_4 - i \chi \tau_4) m_4^0 (f_{\bar{5}5}) \nonumber \\[3mm] 
   & + 2 \left[ 3 \lambda_1 + 2 (2E_4 - i \chi \tau_4) \tau_4 \right] f_{\bar{5}5} + 2 (A_{4\bar{5}} + f_{4\bar{5}} - 2 i \chi A_{\bar{5}4}) f_{45} \,=\, 0.
\label{constraint30}
\end{align}

\subsection*{Algebraic relations}

\be
  A_{5\bar{5}} + A_{\bar{5}5} - 2 E_1 - ( \Lambda + 4 \tau_4^2 ) \chi^2 = 0
\label{algebraic01}
\ee
\be
 f_{5\bar{5}} A_{54} = 0
\label{algebraic02}
\ee

\subsection*{Commutators}
\bea
 \left[ n^0, m_4^0 \right]       &=& - (A_{44} + E_4 \tau_4)\ m_4^0, \label{n0:m40:comm:summary} \\[2mm]
 \left[ n^0, m_5^0 \right]       &=& \lambda_5\ n^0 - \left[ f_{45} + A_{45} + i \chi A_{54} + (E_5 - i \chi \lambda_5) \tau_4 \right]\ m_4^0 
 \notag \\[1mm]
 &\quad& \hspace{45mm}+ \left[ \frac{i \Lambda \chi}{4} - \lambda_1 + \stackrel{5}{\mu}_{\bar{5}1} - (E_4 - i \chi \tau_4) \tau_4 \right] m_5^0, \label{n0:m50:comm:summary} \\[2mm]
 \left[ m_4^0, m_5^0 \right]     &=& \left( \stackrel{5}{\mu}_{\bar{5}4} - \tau_4 \right) m_5^0, \label{m40:m50:comm:summary} \\[2mm]
 \left[ m_5^0, \mbar_5^0 \right] &=& 2 i \chi\ n^0 + 2 i \chi E_4\ m_4^0 + \left( \bar{\lambda}_5 + \stackrel{\bar{5}}{\mu}_{5\bar{5}} \right) m_5^0 -  \left( \lambda_5 + \stackrel{5}{\mu}_{\bar{5}5} \right) \mbar_5^0 \label{m50:m50bar:comm:summary}
\eea

\subsection*{Acknowledgments}

We thank Lode Wylleman for discussions.  This work was supported by the European Research Council grant no.\ ERC-2011-StG 279363-HiDGR. G.B.F.\ is supported by CAPES grant no. 0252/11-5.  M.G.\ is supported by King's College, Cambridge.

\end{document}